\documentclass[twocolumn,tighten]{aastex63}

\usepackage{graphicx}
\usepackage{amsmath}
\usepackage{amssymb}
\usepackage{color}
\usepackage{mathtools}
\usepackage{ulem}
\usepackage[all]{hypcap} 

\usepackage{lineno}

\newcommand\msun{\, {M}_\odot}

\newcommand\kms{\, {\rm km}\,{\rm s}^{-1}}

\newcommand\gpcyr{\, {\rm Gpc}^{-3}\,{\rm yr}^{-1}}

\newcommand\mcl{{M_{\rm CL}}}
\newcommand\rh{{r_{\rm h}}}

\newcommand\zcl{{Z_{\rm CL}}}

\def\apjl{ApJL}%
\def\apjs{ApJS}%
\def\aap{A\&A}%
\begin{document}

\title{Demographics of Hierarchical Black Hole Mergers in Dense Star Clusters}
\shorttitle{Demographics of hierarchical BH mergers}

\author{Giacomo Fragione}
\affil{Department of Physics \& Astronomy, Northwestern University, Evanston, IL 60208, USA}
\affil{Center for Interdisciplinary Exploration \& Research in Astrophysics (CIERA), Northwestern University, Evanston, IL 60208, USA}

\author{Frederic A.~Rasio}
\affil{Department of Physics \& Astronomy, Northwestern University, Evanston, IL 60208, USA}
\affil{Center for Interdisciplinary Exploration \& Research in Astrophysics (CIERA), Northwestern University, Evanston, IL 60208, USA}

\begin{abstract}
With about a hundred mergers of binary black holes (BBHs) detected via gravitational waves by the LIGO-Virgo-KAGRA (LVK) Collaboration, our understanding of the darkest objects in the Universe has taken unparalleled steps forward. While most of the events are expected to consist of BHs directly formed from the collapse of massive stars, some may contain the remnants of previous BBH mergers. In the most massive globular clusters and in nuclear star clusters, successive mergers can produce second- (2G) or higher-generation BHs, and even form intermediate-mass BHs. Overall, we predict that up to $\sim 10\%$, $\sim 1\%$ or $\sim 0.1\%$ of the BBH mergers have one component being a 2G, 3G, or 4G BH, respectively. Assuming that $\sim 500$ BBH mergers will be detected in O4 by LVK, this means that $\sim 50$, $\sim 5$, or $\sim 0.5$ events, respectively, will involve a 2G, 3G, or 4G BH, if most sources are produced dynamically in dense star clusters. With their distinctive signatures of higher masses and spins, such hierarchical mergers offer an unprecedented opportunity to learn about the BH populations in the densest stellar systems and to shed light on the elusive intermediate-mass BHs that may form therein.
\end{abstract}

\section{Introduction}
\label{sect:intro}

The LIGO/Virgo/KAGRA (LVK) Collaboration has recently released the third Gravitational Wave Transient Catalog \citep[GWTC-3, ][]{lvkcoll-O3}, which lists about $80$ confident detections of merging binary black holes (BBHs) detected via gravitational wave (GW) emission. These events are revolutionizing our understanding of compact objects and have made it possible to constrain their masses, spin, and merger rates \citep{lvkcoll-O3-2}.

The origin of these binary mergers is still highly debated. Possible scenarios that could potentially explain BBH mergers include isolated binary star evolution \citep[e.g.,][]{bel16b,demi2016,SperaMapelli2019,BaveraFragos2021}, dynamical formation in globular clusters \citep[e.g.,][]{PortegiesZwartMcMillan2000,askar17,baner18,fragk2018,rod18,sams18,krem2019}, mergers in triple and quadruple systems \citep[e.g.,][]{antoper12,ll18,GrishinPerets2018,arcasedda+2018,fragk2019}, and mergers of compact binaries in galactic nuclei \citep[e.g.,][]{oleary2009,bart17,HoangNaoz2018,LiuLai20199,Tagawa+2020}.

Some of the detected events (such as GW190521, GW190929, and GW190426) are particularly intriguing since one or both components of the merging binary have masses above about $50\msun$. In contrast, stellar evolutionary models predict no BHs with masses larger than about $50\msun$, depending on the progenitor metallicity \citep{woosley2017,limongi2018,bel2020,VinkHiggins2021}, because of the pair-instability process \citep{heger2003,woosley2017}. Since these higher-mass BHs are nevertheless observed, there should exist some astrophysical process that catalyzes their formation. A natural explanation is that BHs more massive than about $50\msun$ are second-generation (2G) BHs, the merger remnants of a previous BBH merger in the core of a dense star cluster \citep[e.g.,][]{gultek2004,antonini2019,frsilk2020,MapelliDall'Amico2021,FragioneKocsis2022,KritosBerti2022}. A fundamental limit for such hierarchical mergers is imposed by the GW recoil kick imparted to merger remnants, which may result in the ejection of the merger remnant if it exceeds the local escape speed \citep[e.g.][]{lou10,lou11}. However, the most-massive globular clusters (GCs) and nuclear star clusters (NSCs) have escape speeds high enough to retain some merger remnants, which can then dynamically assemble into new binaries and merge again via GW emission.

In some cases, repeated mergers could even produce intermediate-mass BHs (IMBHs). IMBHs, with masses between $100\msun$ and $10^5\msun$, represent fundamental building blocks in the cosmological paradigm, but have not been detected beyond any reasonable doubt through either dynamical or accretion signatures \citep[for a review see][]{GreeneStrader2020}. GW detection provides an unparalleled opportunity to survey the sky and detect mergers of IMBHs, making it possible for the first time to constrain their formation, growth, and merger history across cosmic time \citep[e.g.,][]{JaniShoemaker2020,FragioneLoeb2022imbh}. While the current network of GW observatories is still rather limited for BHs with such high masses, the next generation of ground-based observatories and space-based missions promises to detect mergers of IMBH binaries throughout most of the observable Universe.

Simulating hierarchical BH mergers is computationally expensive, and direct $N$-body or Monte Carlo codes cannot currently model the most massive and densest clusters where these events are most frequent \citep[e.g.,][]{Aarseth2003,GierszAskar2019,RodriguezWeatherford2022}. A common approach to tackle the problem has been to use simple order-of-magnitude estimates to assess the rates of 2G or higher-generation BH mergers, which have even led to claims that dense star clusters may produce {\em too many\/} BBH mergers compared to what has been observed by LVK \citep[e.g.,][]{ZevinHolz2022}. In this paper, we use a more realistic semi-analytic framework to model hierarchical mergers in dense star clusters. Our method captures all the essential features of $N$-body and Monte Carlo results for BBH mergers, while allowing us to rapidly sample and access broad regions of the parameter space for even the most massive and densest star clusters. Our results provide for the first time a physically-motivated estimate of the relative fractions of higher-generation mergers as a function of cluster mass and density across cosmic time. We can also then show that some of the specific GW events detected by LVK are consistent with one or both components being a higher-generation BH.

This paper is organized as follows. In Section~\ref{sect:method}, we discuss our semi-analytic method to study hierarchical mergers and the formation of IMBHs. In Section~\ref{sect:results}, we present our results and show that some of the LVK events are consistent with being the result of repeated BBH mergers. Finally, in Section~\ref{sect:concl}, we discuss the implications of our results and draw our conclusions.

\section{Method}
\label{sect:method}

In what follows, we describe the details of the numerical method we use to model the evolution of the BH population in a dense star cluster of mass $\mcl$ and half-mass radius $\rh$.

\subsection{Black holes}
\label{subsec:bh}

We sample stellar masses, $m_*$, from the canonical initial mass function \citep{kro01}
\begin{equation}
\xi(m_*)\propto
\begin{cases}
\left(m_*/0.5\msun\right)^{-1.3}& \text{$0.08\le m_*/\mathrm{M}_\odot\leq 0.50$}\\
\left(m_*/0.5\msun\right)^{-2.3}& \text{$0.50\le m_*/\mathrm{M}_\odot\leq 150$}\,,
\end{cases}
\label{eqn:imf}
\end{equation}
in the range $[20\,\msun,150\,\msun]$, which approximately encompasses the masses of BH progenitors. Given the above IMF, we sample a total of
\begin{equation}
N_{\rm BH} = 3.025 \times 10^{3} \left(\frac{\mcl}{10^6\msun}\right)
\end{equation}
BH progenitors.

We evolve the progenitor mass at a metallicity $\zcl$ using the state-of-the-art version of the stellar evolution code \textsc{sse} \citep{HurleyPols2000}, which includes the most up-to-date prescriptions for stellar winds and remnant formation \citep[see][and references therein]{BanerjeeBelczynski2020}. We do not take into account primordial binaries. After formation, each BH is imparted a natal kick. We calculate BH kicks by sampling from the
same Maxwellian distribution adopted for neutron stars and core-collapse supernovae,
\begin{equation}
    p(v_{\rm natal}) \propto v_{\rm natal}^2\,e^{-v_{\rm natal}^2/\nu^2}\,,
\label{eqn:vnat}
\end{equation}
with 1D velocity dispersion $\nu = 265\kms$ \citep{hobbs2005}, but with BH kicks reduced by a factor $1.4\,\msun/ {m_{\rm BH}}$ assuming momentum conservation \citep{fryerkalo2001}. We check that the natal kicks imparted to the system are below the 3D cluster escape speed
\begin{equation}
    v_{\rm esc}=32\,{\rm km}\,{\rm s}^{-1} \left(\frac{\mcl}{10^5\msun}\right)^{1/2} \left(\frac{\rh}{1\,{\rm pc}}\right)^{-1/2}\, ;
\end{equation}
otherwise we assume the newborn BH to be ejected from the parent cluster. If not ejected from the cluster, the BH sinks to the cluster center over a dynamical friction timescale \citep{Chandrasekhar1943}
\begin{equation}
\tau_{\rm df}\approx 17\ \mathrm{Myr} \left(\frac{20\msun}{m_{\rm BH}}\right)\left(\frac{\mcl}{10^5\msun}\right)^{1/2}\left(\frac{\rh}{1\,\mathrm{pc}}\right)^{3/2}\,.
\label{eqn:tdf}
\end{equation}
We assume that BH natal spins are all zero, consistent with the recent findings of \citet{FullerMa2019}.

\subsection{Cluster evolution}

To model cluster evolution, we follow the elegant approach described in \citet{AntoniniGieles2020,AntoniniGieles2020b}. In this scheme, the cluster is assumed to reach a state of balanced evolution, so that the heat generated by the BBHs in the core and the cluster global properties are related \citep{Henon1961,GielesHeggie2011,BreenHeggie2013}. The cluster energy evolves as
\begin{equation}
    \dot{E}=0.1 \left(\frac{E}{t_{\rm rh}}\right)\,,
\end{equation}
where
\begin{equation}
    E=-0.2\left(\frac{GM_{\rm CL}^2}{\rh}\right)
\end{equation}
is the total energy of the cluster, and
\begin{equation}
    t_{\rm rh}=\frac{0.138}{\langle m\rangle \psi \ln \Lambda}\left(\frac{\mcl r_{\rm h}^3}{G}\right)^{1/2}
\end{equation}
is the average relaxation time. In the previous equation $\langle m\rangle\approx 0.6\msun$ is the mean stellar mass in the cluster and $\ln \Lambda=10$ is the Coulomb logarithm. The quantity $\psi$ depends on the stellar mass function within the cluster half-mass radius; $\psi=1$ for systems with objects of all equal masses, but it can be between 1.5 and 2 for a realistic mass spectrum \citep{SpitzerHart1971a,SpitzerHart1971b}. To account for the role of BHs, we parameterize it as \citep{AntoniniGieles2020}
\begin{equation}
    \psi=1+1.47\left(\frac{M_{\rm BH}/\mcl}{0.01}\right)\,,
\end{equation}
where $M_{\rm BH}$ is the total mass in BHs. The balanced evolution starts at a time \citep{AntoniniGieles2020}
\begin{equation}
    t_{\rm cc}= 3.21\,t_{\rm rh,0}\,,
\end{equation}
where $t_{\rm rh,0}$ is the initial relaxation time.

The star cluster is considered isolated, thus we neglect the effect of any galactic tidal fields, and loses mass as a result of mass loss from stars ($\dot{M}_{\rm sev}$), evaporation ($\dot{M}_{\rm ev}$), and BH ejections ($\dot{M}_{\rm BH}$)
\begin{equation}
    \dot{M}_{\rm CL}=\dot{M}_{\rm sev}+\dot{M}_{\rm ev}+\dot{M}_{\rm BH}\,.
\end{equation}
We parameterize the mass loss from stars as \citep{AntoniniGieles2020}
\begin{equation}
    \dot{M}_{\rm sev}=
    \begin{cases}
    0 & t<2\,{\rm Myr},\\
    -8.23\times 10^{-2} (M_* /t) & t\ge 2\,{\rm Myr}\,,
    \end{cases}
    \label{eqn:msev}
\end{equation}
while cluster evaporation is calculated as \citep{gne14}
\begin{equation}
    \dot{M}_{\rm ev} = 1.17\times 10^4\,M_\odot \,{\rm Gyr}^{-1}\,;
    \label{eqn:mev}
\end{equation}
for mass loss resulting from BH ejections we refer to the next subsection. The cluster radius expands adiabatically as a result of stellar evolution \citep{AntoniniGieles2020}
\begin{equation}
    \dot{r}_{\rm h,sev} = -\frac{\dot{M}_{\rm sev}}{\mcl}\rh\,,
\end{equation}
and as a result of balanced evolution and relaxation
\begin{equation}
    \dot{r}_{\rm h,rlx}=\zeta\frac{\rh}{t_{\rm rh}}+2\frac{\dot{M}_{\rm CL}}{\mcl}\rh\,;
\end{equation}
therefore
\begin{equation}
    \dot{r}_{\rm h}=
    \begin{cases}
    \dot{r}_{\rm h,sev} & t<t_{\rm cc},\\
    \dot{r}_{\rm h,sev}+\dot{r}_{\rm h,rlx} & t\ge t_{\rm cc}\,.
    \end{cases}
    \label{eqn:rh}
\end{equation}

\begin{figure*} 
\centering
\includegraphics[scale=0.485]{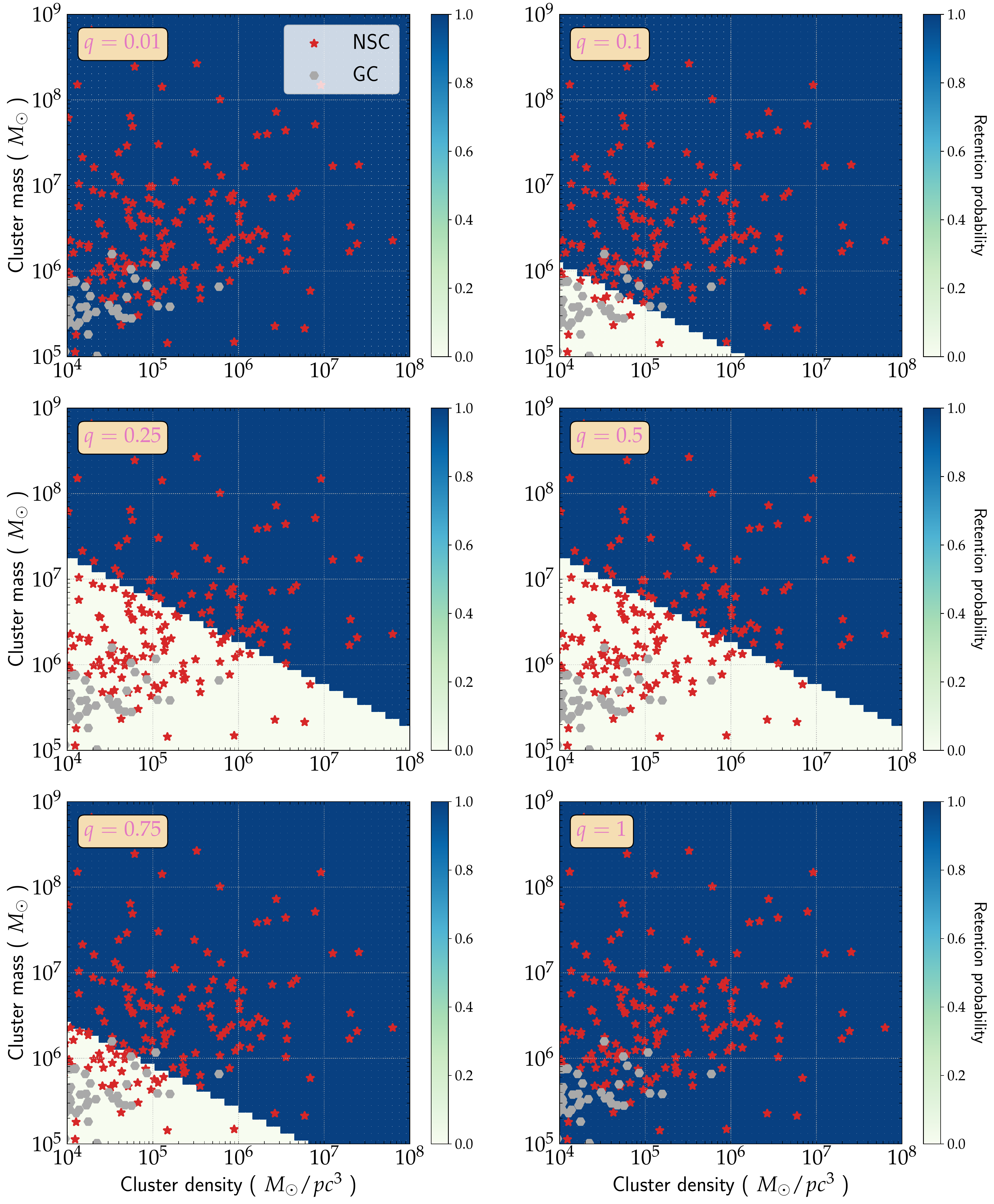}
\caption{Probability to retain the merger remnant of a BBH as a function of the host cluster mass and density for different values of the binary mass ratio, from $q=0.01$ (top-left panel) to equal masses (bottom-right panel). Both BHs in the binary are assumed to be from a first generation, with initial spins $\chi_1=\chi_2=0$ (cf. Figures.~\ref{fig:ret2} and~\ref{fig:ret3}). Gray hexagons represent Milky Way globular clusters from \citet{BaumgardtHilker2018}, while red stars represent nuclear star clusters from \citet{georg2016}.}
\label{fig:ret1}
\end{figure*}

\begin{figure*} 
\centering
\includegraphics[scale=0.485]{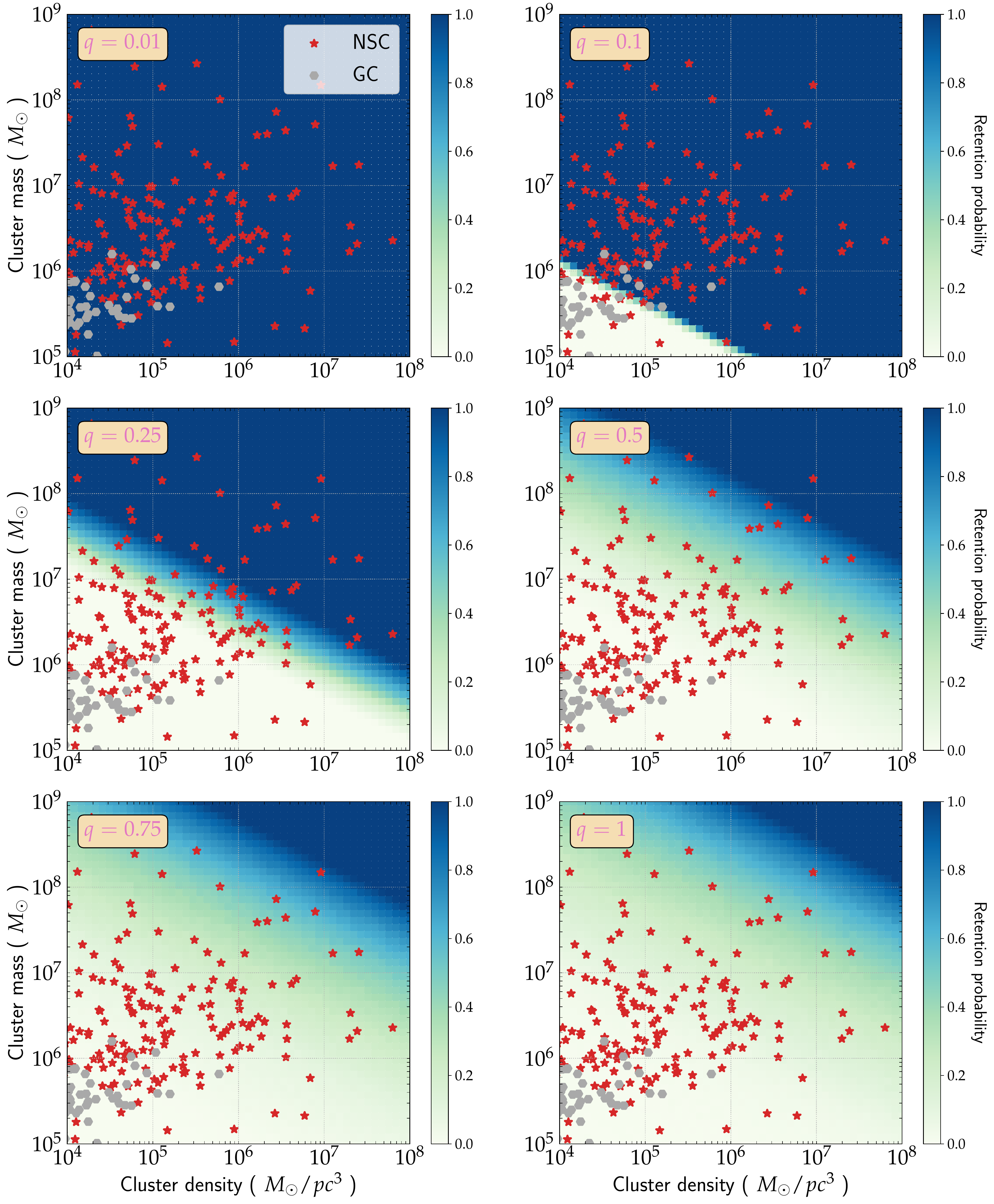}
\caption{Same as Figure~\ref{fig:ret1}, but here for binaries containing one first-generation BH and one second-generation BH, with spins $\chi_1=0$ and $\chi_2=0.7$, respectively.}
\label{fig:ret2}
\end{figure*}

\begin{figure*} 
\centering
\includegraphics[scale=0.485]{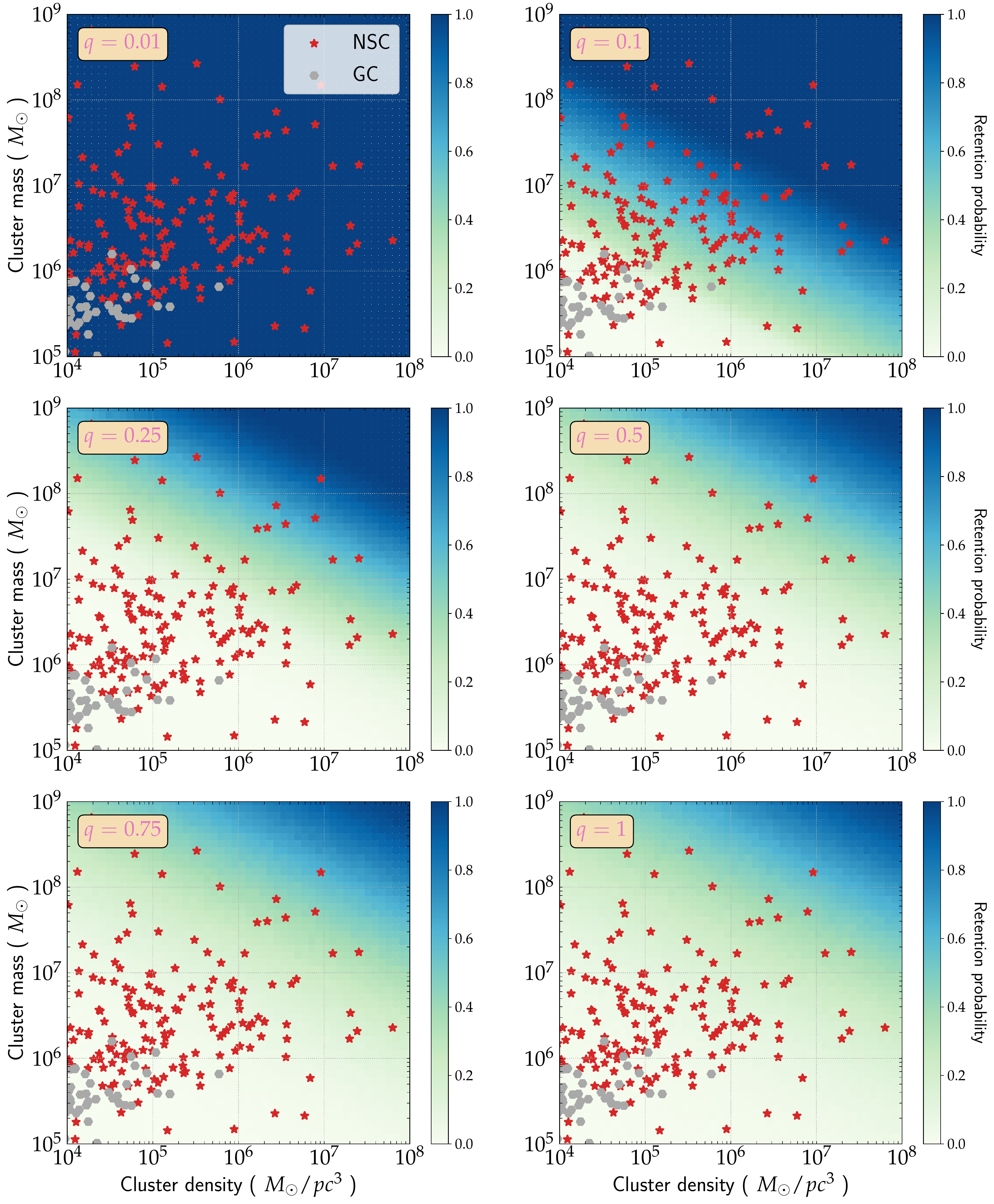}
\caption{Same as Figure~\ref{fig:ret1}, but here for binaries containing two second-generation BHs, with spins $\chi_1=\chi_2=0.7$.}
\label{fig:ret3}
\end{figure*}

\subsection{Binary black hole mergers}

Balanced evolution imposes that the required heating rate of the cluster is balanced with the loss of energy from the BBHs in its core. Assuming that one BBH (of component masses $m_1$ and $m_2$) dominates the heating at all times, we require that $\dot{E}_{\rm bin}(t)=-\dot{E}(t)$, where $\dot{E}_{\rm bin}(t)$ is the rate of energy loss from the binary \citep{antonini2019}.

The initial population of (first-generation) BHs is obtained directly from the evolution of the massive stars we sample in the star cluster, as described in Section~\ref{subsec:bh}. We sample the masses of the binary (that we assume dominates the heating at all times) considering that the for 3-body binary formation the likelihood of forming a BH binary with component masses $m_1$ and $m_2$ is $\propto (m_1+m_2)^5$ \citep{mors2015}. We take its initial semi-major axis to be at the hard-soft boundary \citep{Heggie1975}
\begin{equation}
    a_{\rm bin,ini} = \frac{Gm_1 m_2}{\langle m \rangle v_{\rm disp}^2}\,,
\end{equation}
where $v_{\rm disp}=0.2\,v_{\rm esc}$, as appropriate for a King model with initial moderate concentration \citep{King1962}.

We assume that every binary-single interaction in the cluster core leads to a decrease in the semi-major axis of the binary we form, until the binary evolution becomes eventually dominated by GW energy loss. As a consequence, the binary semi-major axis will decrease after each interaction as
\begin{equation}
    \frac{\Delta a_{\rm bin}}{a_{\rm bin}}=\delta-1\,,
\end{equation}
with $\delta=7/9$ for equal masses \citep{quin1996,HeggieHut2003,SamsingMacLeod2014}, which we generalize to 
\begin{equation}
    \delta = 1.0 - \frac{2}{3}\left(\frac{m_3}{m_1+m_2+m_3}\right)\,,
    \label{eqn:deltabh}
\end{equation}
where $m_3$ is the mass of the single BH that interacts with the target binary. We sample $m_3$ considering that the interaction probability is $\propto m_3$ \citep{AntoniniGieles2022}. Therefore, the timescale during which the binary-single interaction occurs can be estimated as
\begin{equation}
    \Delta t_{i}=\left(\frac{1}{\delta}-1\right) \frac{Gm_1m_2}{2a_{\rm bin}\dot{E}_{\rm bin}}\,.
\end{equation}
When repeated over several binary-single interactions, the overall timescale to transition to the GW-dominated regime is
\begin{equation}
    \tau = \sum_i \Delta t_{i}\,.
\end{equation}
We assume that during each binary-single encounter, the binary receives a large angular momentum kick such that the phase space is stochastically explored and uniformly covered by the periapsis values \citep[e.g.,][]{KatzDong2012}. The transition to the GW-dominated regime happens whenever the BBH eccentricity, drawn from a thermal distribution at each scattering (``in-cluster merger'')
\begin{equation}
    e_{\rm bin}>\left[1- 1.3\left(\frac{G^4 m_1^2 m_2^2 (m_1+m_2)}{c^5 \dot{E}_{\rm bin}}\right)^{1/7}a_{\rm bin}^{-5/7} \right]^{1/2}\,.
\end{equation}
However, the sequence of binary-single scatterings can be halted if either the binary mergers in cluster before the following interaction or if the binary is ejected. In the first case, we divide each binary-single encounter in a set of $20$ resonant intermediate states, and we assume that the binary eccentricity after each state is sampled from a thermal distribution \citep{Samsing2018}. A merger (``GW capture'') occurs before the next state if \citep{FragioneLoebkr2020}
\begin{equation}
    e_{\rm bin, int} > 1 - 1.6 \left(\frac{R_{\rm S, 1}}{a_{\rm bin}}\right)^{5/7} q^{2/7} (1+q)^{1/7}\,,
\end{equation}
where $q=m_2/m_1$ and $R_{\rm S, 1}$ is the Schwarzshild radius of the primary BH. For what concerns ejections, during a binary-single encounter the binary receives a recoil kick \citep{antoras2016}
\begin{equation}
    v_{\rm 12}=\left(\frac{1}{\delta}-1\right)\frac{G\mu_{12} m_3}{(m_1+m_2+m_3)a_{\rm bin}}\,,
\end{equation}
where $\mu_{12}=m_1m_2/(m_1+m_2)$, and the third BH a recoil kick
\begin{equation}
    v_{\rm 3}=\frac{m_1+m_2}{m_3} v_{\rm 12}\,,
\end{equation}
as a result of energy and momentum conservation. If $v_{\rm 12}>v_{\rm esc}$, the binary is ejected from the parent cluster and may eventually merge via GW emission in the field (``ejected merger''). If $v_{\rm 3}>v_{\rm esc}$, the third BH of mass $m_3$ is assumed to be ejected from the host cluster. We model the mass lost by the cluster in BHs, $\dot{M}_{\rm BH}$, as the sum of all the BHs ejected (binaries and singles) during three-body interactions. Note that we self-consistently keep track of the masses, spins, and generations of each BH within its host star cluster. After the BBH either merges or is ejected from the cluster, we form a new BBH using the updated BH population, as described at the beginning of this section.

\subsection{Recoil kicks and merger remnants}

As a result of the anisotropic emission of GWs at merger, the merger remnant is imparted a recoil kick that depends on the asymmetric mass ratio $\eta=q/(1+q)^2$ and on the magnitude of the dimensionless spin parameters, $\chi_1$ and $\chi_2$. In our models, spin orientations are assumed to be isotropic, as appropriate for merging binaries assembled dynamically. We model the recoil kick as \citep{lou10,lou12}
\begin{equation}
\textbf{v}_{\mathrm{kick}}=v_m \hat{\textbf{e}}_{\perp,1}+v_{\perp}(\cos \xi \hat{\textbf{e}}_{\perp,1}+\sin \xi \hat{\textbf{e}}_{\perp,2})+v_{\parallel} \hat{\textbf{e}}_{\parallel}\,,
\label{eqn:vkick}
\end{equation}
where
\begin{eqnarray}
v_m&=&A\eta^2\sqrt{1-4\eta}(1+B\eta)\\
v_{\perp}&=&\frac{H\eta^2}{1+q}(\chi_{2,\parallel}-q\chi_{1,\parallel})\\
v_{\parallel}&=&\frac{16\eta^2}{1+q}[V_{1,1}+V_A \tilde{S}_{\parallel}+V_B \tilde{S}^2_{\parallel}+V_C \tilde{S}_{\parallel}^3]\times \nonumber\\
&\times & |\mathbf{\chi}_{2,\perp}-q\mathbf{\chi}_{1,\perp}| \cos(\phi_{\Delta}-\phi_{1})\,.
\end{eqnarray}
The $\perp$ and $\parallel$ refer to the direction perpendicular and parallel to the orbital angular momentum, respectively, while $\hat{e}_{\parallel, 1}$ and $\hat{e}_{\parallel, 2}$ are orthogonal unit vectors in the orbital plane. We have also defined the vector
\begin{equation}
\tilde{\mathbf{S}}=2\frac{\boldsymbol{\chi}_{2}+q^2\boldsymbol{\chi}_{1}}{(1+q)^2}\,,
\end{equation}
$\phi_{1}$ as the phase angle of the binary, and $\phi_{\Delta}$ as the angle between the in-plane component of the vector
\begin{equation}
\boldsymbol{\Delta}=M^2\frac{\boldsymbol{\chi}_{2}-q\boldsymbol{\chi}_{1}}{1+q}
\end{equation}
and the infall direction at merger. Finally, we adopt $A=1.2\times 10^4$ km s$^{-1}$, $H=6.9\times 10^3$ km s$^{-1}$, $B=-0.93$, $\xi=145^{\circ}$ \citep{gon07,lou08}, and $V_{1,1}=3678$ km s$^{-1}$, $V_A=2481$ km s$^{-1}$, $V_B=1793$ km s$^{-1}$, $V_C=1507$ km s$^{-1}$ \citep{lou12}. We adjust the final total mass and spin of the merger remnant using the results of \citet{Jimenez-FortezaKeitel2017}, which we generalized to precessing spins following the approach in \citet{HofmannBarausse2016}. 

Whenever $v_{\rm kick}>v_{\rm esc}$, the remnant is ejected from the host cluster; otherwise, it sinks back to the cluster core on the dynamical friction timescale (see Eq.~\ref{eqn:tdf}). In our simulations, we keep track of the masses, spins, and generations of each BH that is retained within its host cluster.

\subsection{Growth of intermediate-mass black holes}

If successfully retained, a remnant BH may eventually keep merging and grow into an IMBH. Whenever its mass is sufficiently large, the interaction between a binary composed of a stellar-mass BH and IMBH with a third stellar-mass BH may change characteristics compared to what previously described, to eventually transition to a behaviour similar to the case of supermassive BH binaries in galactic nuclei. While the amount of energy subtracted per encounter is still small and likely approximately described by Eq.~\ref{eqn:deltabh}, the binary is not going to explore uniformly the eccentricity space; rather the eccentricity increases as a function of time
\begin{equation}
    \Delta e_{\rm bin} = \kappa \frac{\Delta a_{\rm bin}}{a_{\rm bin}}\,,
\end{equation}
with $\kappa=0.01$ \citep[e.g.,][]{quin1996,sesa2006}. Note that, however, \citet{BonettiRasskazov2020} showed that for mass ratios $\lesssim 10^{-3}$ the eccentricity growth rate may become negative on average, due to a subset of interacting stars captured in meta-stable counter-rotating orbits, which tend to inject angular momentum from the binary. We switch our eccentricity prescription whenever the primary mass in the merging binary is larger than $1000\msun$.

\section{Hierarchical mergers}
\label{sect:results}

In this Section, we study how different generations of BHs contribute to the overall population of detected mergers and we compare their properties with those of LVK-detected BBHs. For a comparison of our models with results from Monte Carlo simulations using the \textsc{cmc} code, see the Appendix.

\begin{figure*} 
\centering
\includegraphics[scale=0.48]{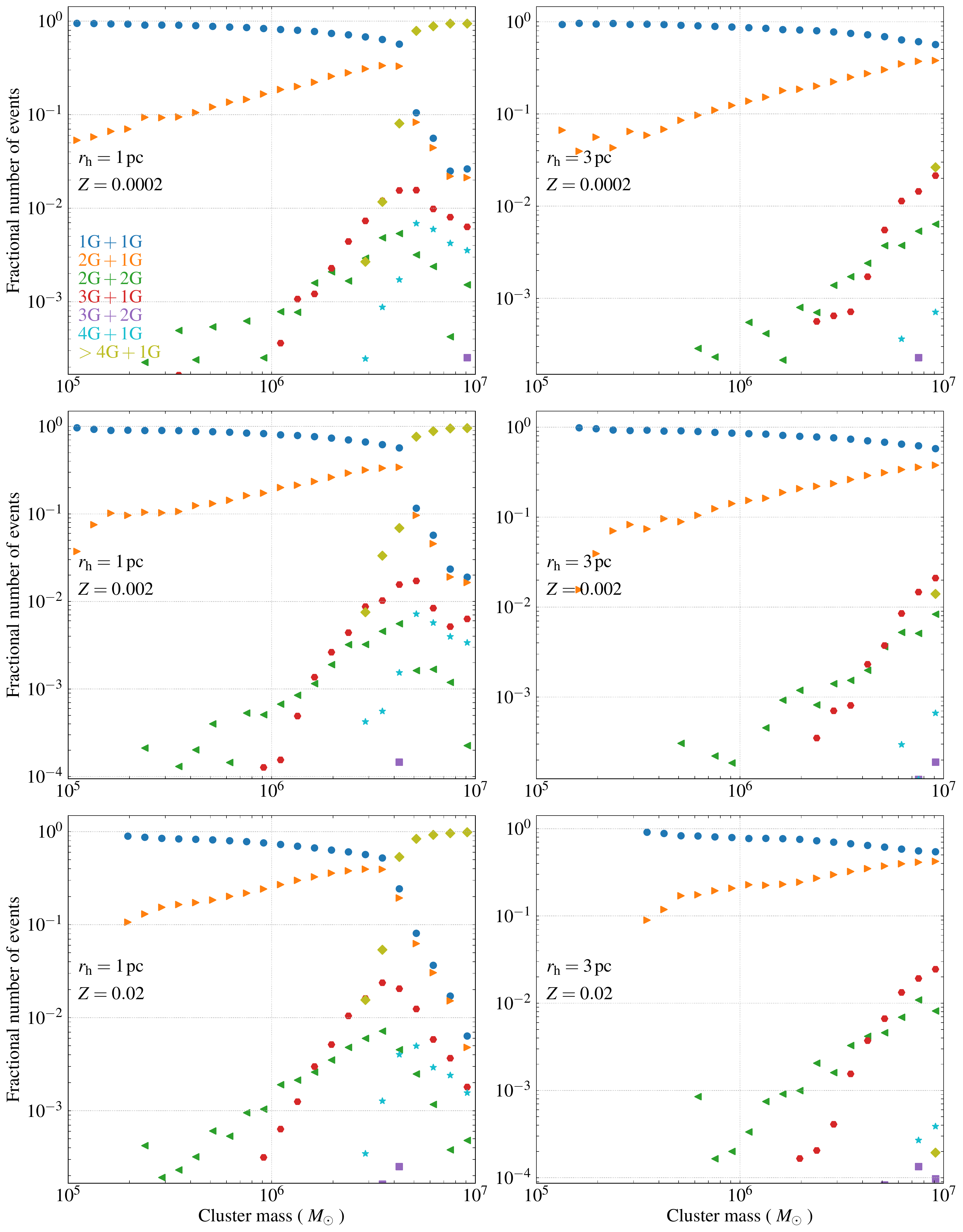}
\caption{Fractional number of events for different generations as a function of the host cluster mass and for $r_h=1\,$pc (left) and $r_h=3\,$pc (right). Top panel: $Z=0.0002$; central panel $Z=0.002$; bottom panel $Z=0.02$.}
\label{fig:gener}
\end{figure*}

We start by discussing the likelihood of retaining the remnant of a BBH merger in a dense star cluster as it is imparted a recoil kick through anisotropic emission of GWs. We first consider the case where both BHs in the merging binary are from the first generation, which we assume to be non-spinning BHs (as expected based on recent models of stellar evolution; see  \citet{FullerMa2019}). Figure~\ref{fig:ret1} shows the probability to retain the merger remnant as a function of the host cluster mass and density, and for different values of the mass ratio. We also plot the mass and half-mass density of Milky Way's globular clusters from \citet{BaumgardtHilker2018} and of nuclear star clusters from \citet{georg2016}. In case of non-spinning BHs, the recoil kick is always very low in the case of very low mass ratios, or even vanishes for equal masses. Therefore, the remnant 2G BH is always retained within its parent cluster. For intermediate mass ratios, however, the retention likelihood significantly decreases. In Figure~\ref{fig:ret2}, we show the retention probability in the case one of the two BHs in the binary is of a second generation. In this case, the 2G BH has a spin of about $0.7$, considering that its progenitors were not spinning \citep[e.g.,][]{buo08}. Since introducing a spin adds asymmetry in the emission of GWs, the likelihood of retaining the remnant decreases with respect to the previous case. The retention probability decreases further in the case both BHs are of a second generation, as illustrated in Figure~\ref{fig:ret3}. It is clear that only the most massive and dense clusters could form, and eventually produce mergers of, BHs beyond the second generation, with 3G BHs more likely to come from the 2G+1G merger channel, rather than the 2G+2G channel.

Figure~\ref{fig:gener} shows the fractional number of events for different generations as a function of the host cluster mass and metallicity, in the case of $r_h=1\,$pc (left) and $r_h=3\,$pc (right). First, note that the overall trends mainly depend on the initial cluster mass and half-mass radius, and not on its metallicity. Second, as expected, the denser the system is, the more likely it is to produce mergers of BBH of a higher generation. For $r_{\rm h}=1\,$pc, we find that 1G+1G mergers represent most of the population of BBH mergers for clusters masses $\sim 10^5\msun$. The contribution of 1G+1G mergers decreases at higher masses, with 2G+1G mergers becoming $\sim 10\%$ of the population for clusters masses $\sim 10^6\msun$, up to about $30\%$ for clusters of $\sim 5\times 10^6\msun$, before decreasing in importance in favor of higher-generation mergers. Mergers of 3G+1G BBHs start happening for clusters masses above $\sim 10^6\msun$ and are typically never more than $\sim 1\%$ of the mergers, while 4G+1G mergers are assembled only for cluster masses $\sim 3\times 10^6\msun$.

This trend is then essentially reproduced by any higher-generation merger (5G+1G, 6G+1G, and so on). The reason is that the mass ratio of the merger is now small enough that the recoil kick imparted to the remnant is not large enough to eject it from the parent cluster. Moreover, the spin of the remnant decreases on average \citep{FragioneKocsis2022}, further suppressing the recoil kick. At this point, this growing BH is massive enough to dominate the BBH mergers and eventually it grows to form an IMBH. This is clearly shown by the fact that higher-generation mergers (``$>$4G+1G'' points in Figure~\ref{fig:gener}) represent essentially most of the events at high cluster masses.

\begin{figure} 
\centering
\includegraphics[scale=0.6]{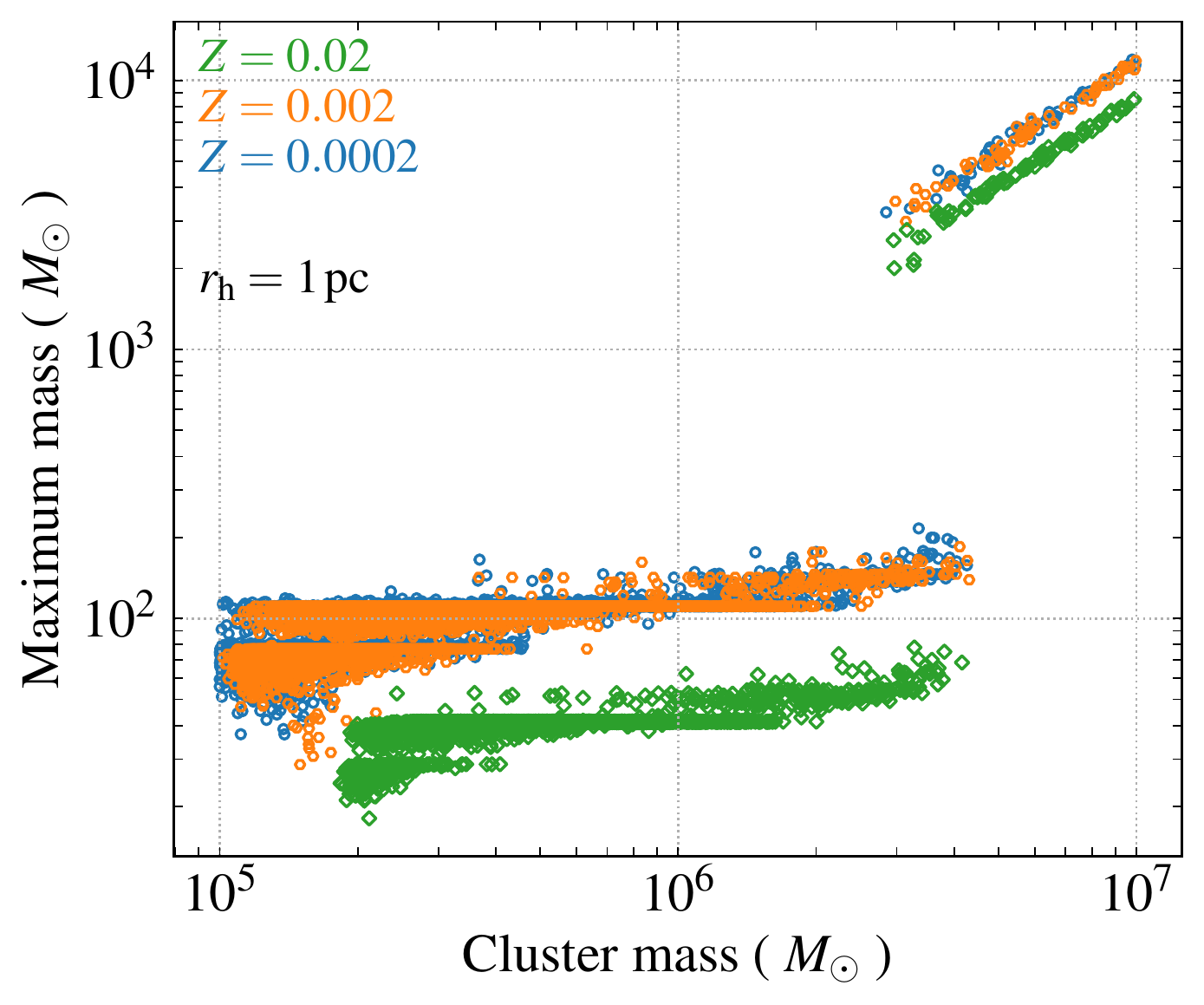}
\caption{Maximum BH mass formed via repeated mergers as a function of the cluster mass. A transition to IMBH formation is seen at around $4\times 10^6\msun$. The half-mass radius of all clusters here is fixed at $1\,$pc. For models with a half-mass radius of $3\,$pc, a similar transition occurs around $10^7\msun$.}
\label{fig:mmax}
\end{figure}

\begin{figure} 
\centering
\includegraphics[scale=0.6]{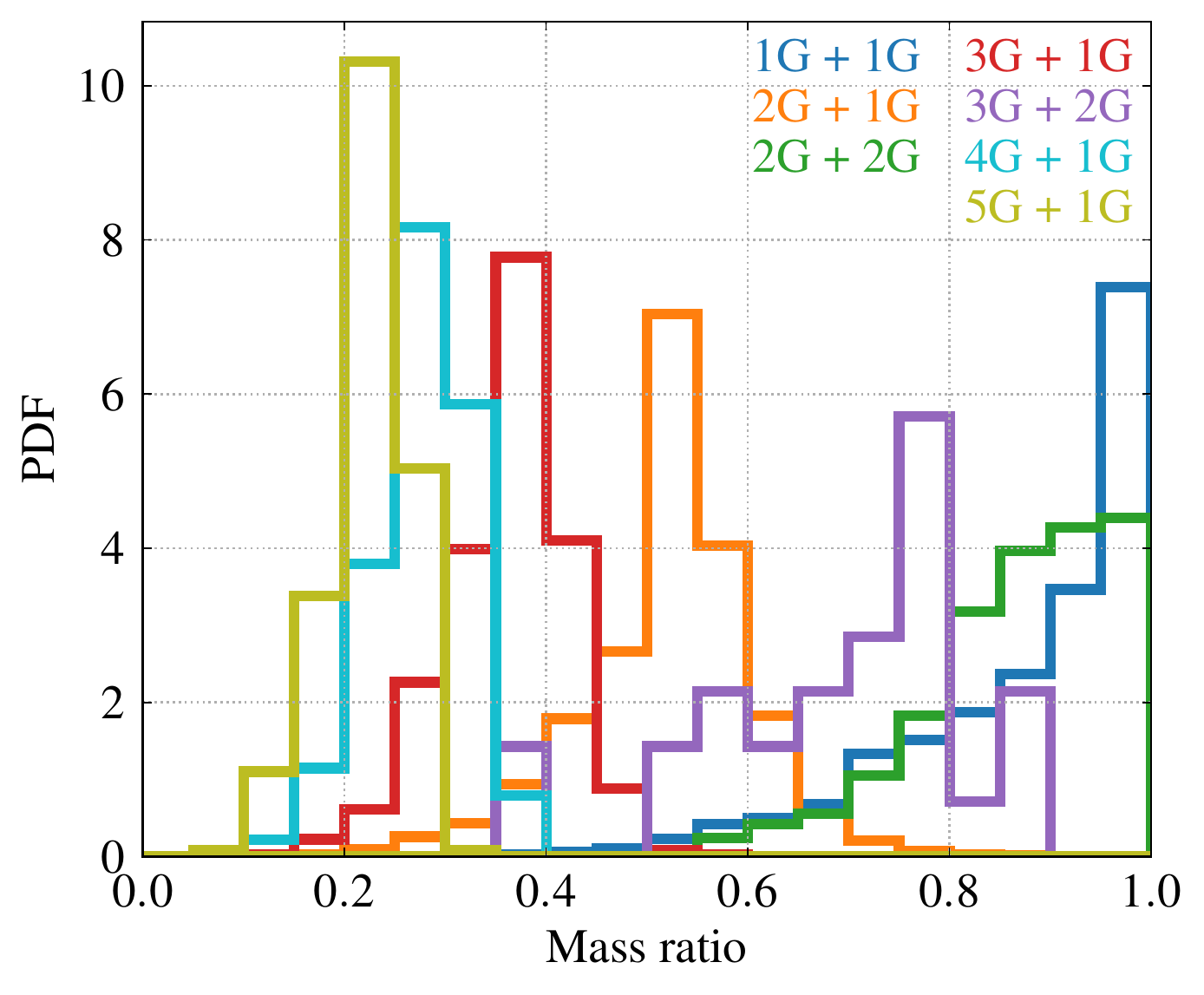}
\caption{Probability distribution function of mass ratios for merging BBHs of different generations.}
\label{fig:massratio}
\end{figure}

This is also illustrated in Figure~\ref{fig:mmax}, where we plot the maximum BH mass produced via hierarchical mergers as a function of the cluster mass, assuming $r_{\rm h}=1\,$pc. It is clear that there is a transition around $4\times 10^6\msun$, after which a single BH dominates the mergers and can grow up to the IMBH regime, $\gtrsim 1000\msun$. This trend does not depend on the metallicity of the cluster, with higher metallicities simply translating into a lower mass of the final IMBH, as a result of the lower initial stellar BH masses. Indeed, a cluster born with solar metallicity can produce BHs with masses just up to about $15\msun$, unlike clusters born at low metallicities, whose BHs at birth can be as massive as about $50\msun$ BHs \citep[e.g.,][]{BanerjeeBelczynski2020}. It is important to note that including 2G+2G and 3G+2G is crucial to characterize the transition to dense star clusters that can eventually form an IMBH. Indeed, the recoil kick imparted to the remnants of 2G+2G and 3G+2G mergers could be significantly larger than the case of 2G+1G and 3G+1G mergers, respectively, where the secondary BH is of a first generation. Therefore, accounting for 2G+2G and 3G+2G mergers is critical in determining if an IMBH could be formed through hierarchical mergers, and even the most massive and dense star clusters in the Universe have only a small likelihood to succeed in this process (see also Figures~\ref{fig:ret2}-\ref{fig:ret3}). For models with a half-mass radius of $3\,$pc, a similar transition occurs around $10^7\msun$.

Among mergers where both components are of a second or higher generation, we find that 2G+2G mergers never represent more than $\sim 0.1\%$ and $\sim 1\%$ of the merging BBH population in star clusters with masses $\sim 10^5\msun$ and $\sim 10^6\msun$, while 3G+2G mergers can only account for $\lesssim 0.1\%$ of the overall population and are assembled only in star clusters with masses $\gtrsim 5\times 10^6\msun$.

We find similar overall trends in the case star clusters have half-mass radius of $r_{\rm h}=3$\,pc, but shifted towards higher cluster masses. Indeed, these clusters are less dense than the case of $r_{\rm h}=1$\,pc, thus a higher cluster mass is needed in order to retain and catalyze the mergers of BBH of a higher generation.

We report in Figure~\ref{fig:massratio} the mass ratio distribution for different generations. The peak of the 1G+1G and 2G+2G mergers is around unity, as result of the fact that the dynamical encounters in the core of dense star clusters tend to process and catalyze the merger of BHs of comparable masses. Then, each generation has a distinctive distribution, with a peak that depends on which generation the two merging BHs belong to. For example, the mass ratio distribution of a merger of a 3G BH and 2G BH is going to be peaked around $2/3$, and so no. Therefore, we find that the mass ratio distribution of 2G+1G, 3G+1G, 3G+2G, 4G+1G, 5G+1G mergers is peaked at about $0.5$, $0.33$, $0.75$, $0.25$, $0.2$, respectively. 

\begin{figure} 
\centering
\includegraphics[scale=0.575]{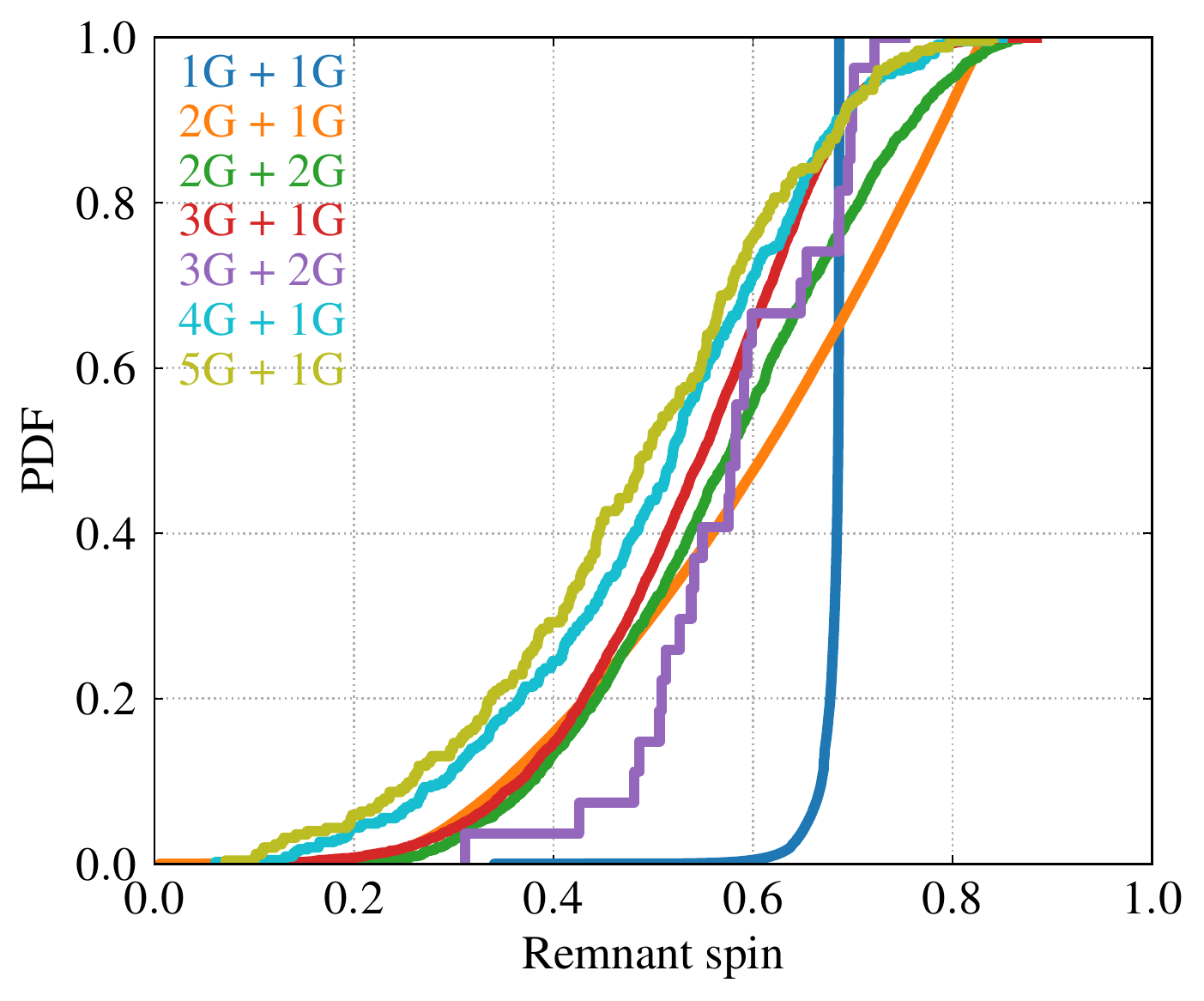}
\includegraphics[scale=0.575]{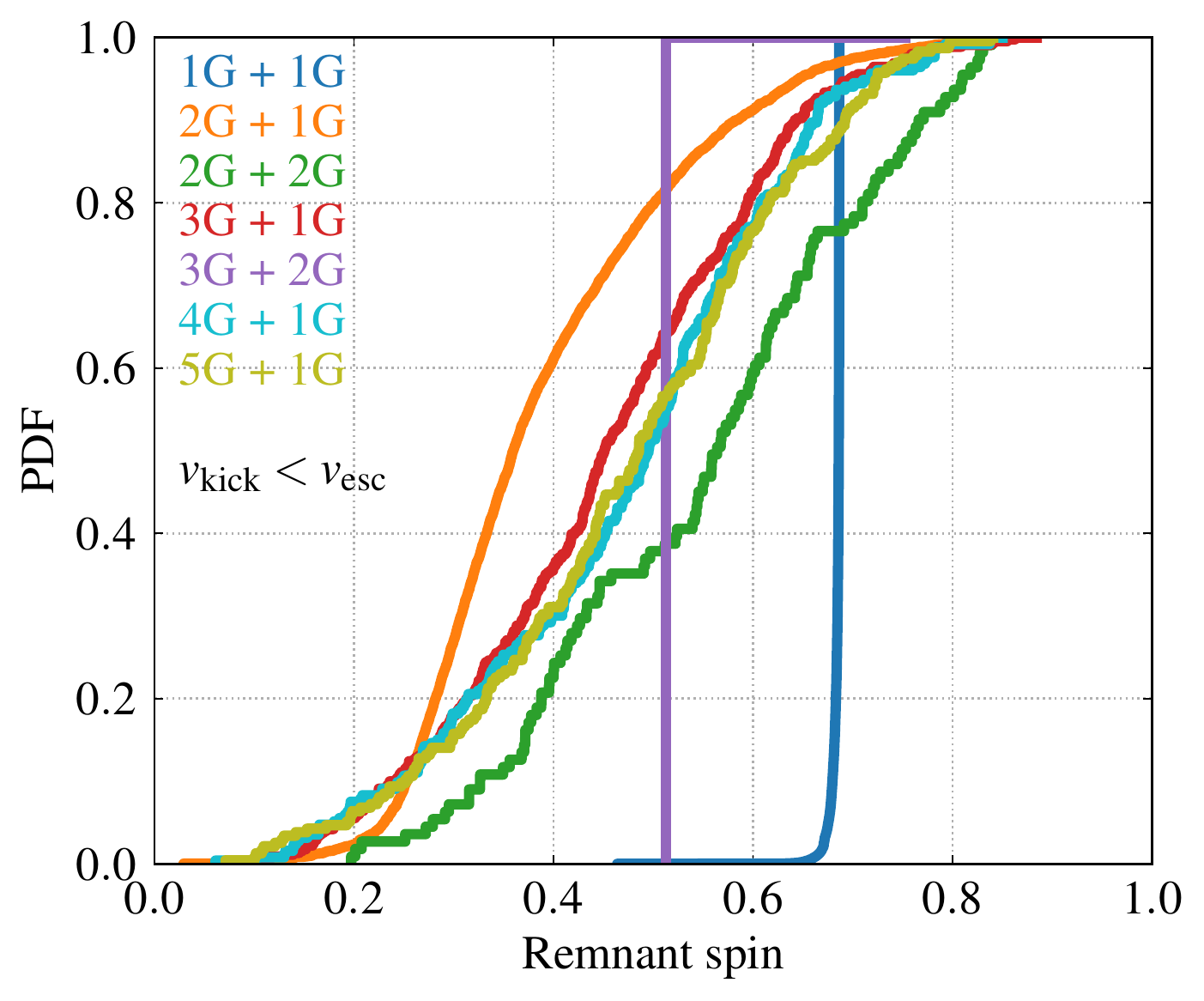}
\includegraphics[scale=0.575]{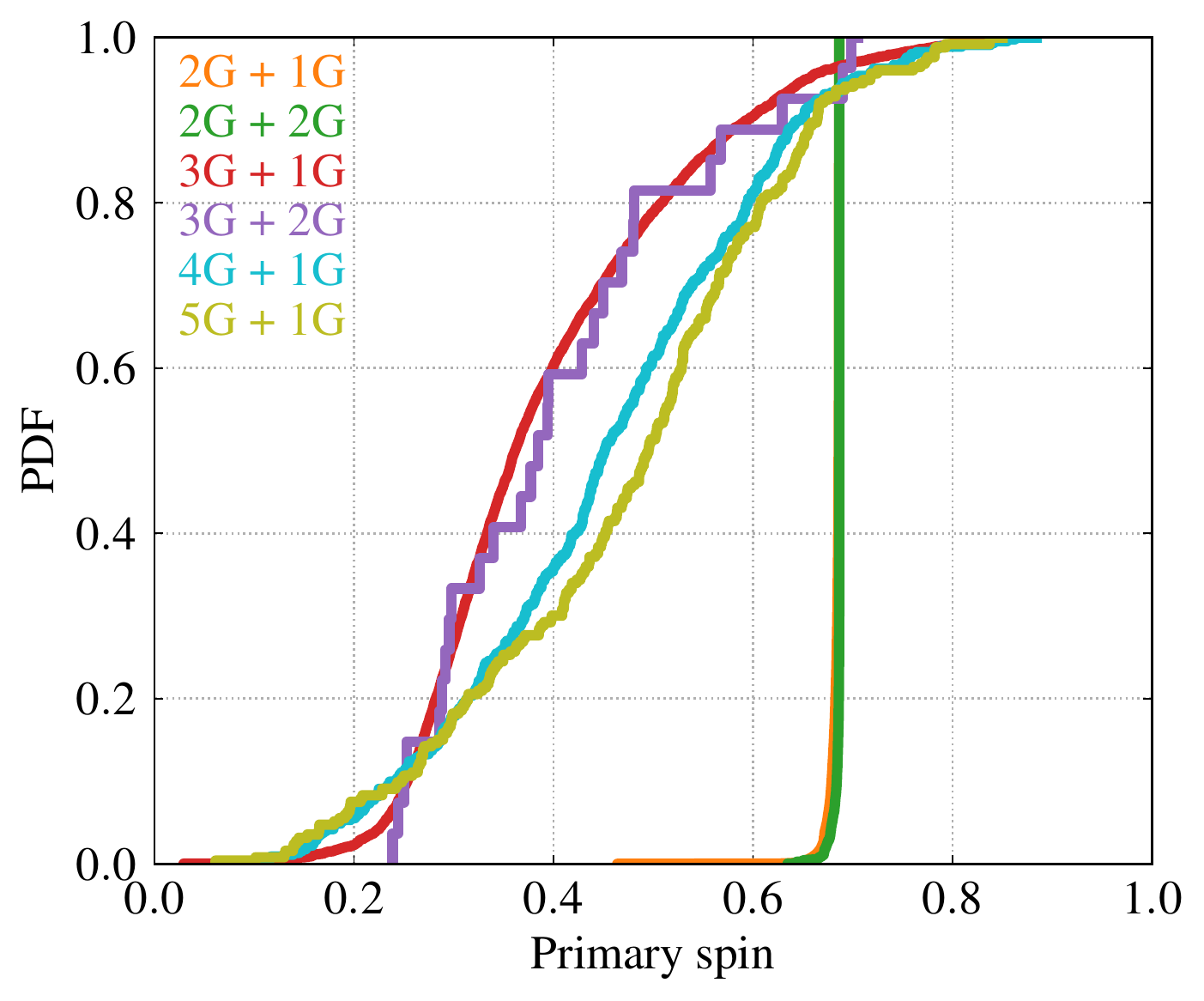}
\caption{Cumulative distribution function of the dimensionless spin magnitude for all merger remnant BHs (top panel), for remnant BHs that are retained within their parent cluster (central panel), and for the primaries of BBHs that merge (bottom panel).}
\label{fig:spin}
\end{figure}

Figure~\ref{fig:spin} shows the cumulative distribution function of the spin of the remnant BHs after a a BBH merger, the spin of the remnant BHs that are retained within their parent cluster, and the spin of the primary masses of the BBHs that merge. These plots show a quite general picture, with the first merger producing a remnant with a spin parameter of about $0.7$ starting from two slowly spinning BHs, which then tends to decrease with subsequent mergers, eventually producing a negative correlation between mass and spin \citep[e.g.,][]{antonini2019,FragioneKocsis2022}. The reason is that the final inspiral and deposition of angular momentum happen at random angles with respect to the spin of the more massive BH, assuming an isotropic geometry of BBH mergers as appropriate to a dynamical environment. The growing BH undergoes a damped random walk in the evolution of its spin because retrograde orbits become unstable at a larger specific angular momentum than do prograde orbits, so it is easier to decrease than to increase the spin magnitude, ending up having a spin of about $0.3$ by the time it reaches $\sim 1000\msun$. It is interesting noting that, while the spin of a 3G BH is around $0.6$, the ones that are retained (coming mostly from a 2G+1G merger) have an average spin of about $0.3$.

We now proceed with computing the merger rates for different generations of BBH mergers. We compute the rates as
\begin{eqnarray}
    R(z) & = & K \frac{d}{dt_{\rm lb}} \int \int \int \int dM_{\rm CL}\,dr_{\rm h}\,dZ\,dz_{\rm f}\, \frac{dt_{\rm lb}}{dz_{\rm f}} \times \nonumber \\
    & \times & \frac{\partial N_{\rm events}}{\partial M_{\rm CL}\,\partial r_{\rm h}\,\partial Z\,\partial z_{\rm f}}\,\Psi(M_{\rm CL}, r_{\rm h}, Z, z_{\rm f})\,,
    \label{eqn:rates}
\end{eqnarray}
where $N_{\rm events}$ is the number of events, $t_{\rm lb}$ is the look-back time at redshift $z$\footnote{For our calculations we assume the cosmological parameters from Planck 2015 \citep{PlanckCollaborationAde2016}.}, and $\Psi(M_{\rm CL}, r_{\rm h}, Z, z_{\rm f})$ is a weighting function that accounts for the cosmic distribution of cluster masses, sizes, metallicities, and formation times. Cluster masses are weighted proportionally to $M_{\rm CL}^{-2}$ up to $M_{\rm CL}^{\max}=10^7\msun$, while their formation times are assumed proportional to $\exp\left[-(z-z_{\rm f})^2 / (2\sigma_{\rm f}^2)\right]$, with $z_{\rm f}=3.2$ and $\sigma_{\rm f}=1.5$ \citep{MapelliDall'Amico2021} and normalized such that the cluster density is $2.5\,{\rm Mpc}^{-3}$ in the local Universe \citep[e.g.,][]{PortegiesZwartMcMillan2010}. Metallicities are sampled from a log-normal distribution with mean given by \citep{MadauFragos2017}
\begin{equation}
\log \langle Z/{\rm Z}_\odot \rangle = 0.153 - 0.074\,z^{1.34}
\end{equation}
and a standard deviation of $0.5$~dex. Finally, $K$ in Equ.~\ref{eqn:rates} is a correction factor that accounts for the evolution of the cluster density from cluster formation times to present day. We take $K=32.5^{+86.9}_{-17.7}$ as found in the analysis of \citet{AntoniniGieles2020}, which is also consistent with the inferred value needed to reproduce the LVK rate of dynamical mergers \citep{FishbachFragione2023}. For initial cluster sizes, we simply consider the two cases where all star clusters are born with half-mass radius $r_{\rm h}=1\,$pc, or all star clusters are born with half-mass radius $r_{\rm h}=3\,$pc; these represent the typical spread of observed values for young clusters in the local Universe \citep[e.g.,][]{PortegiesZwartMcMillan2010}.

\begin{figure} 
\centering
\includegraphics[scale=0.6]{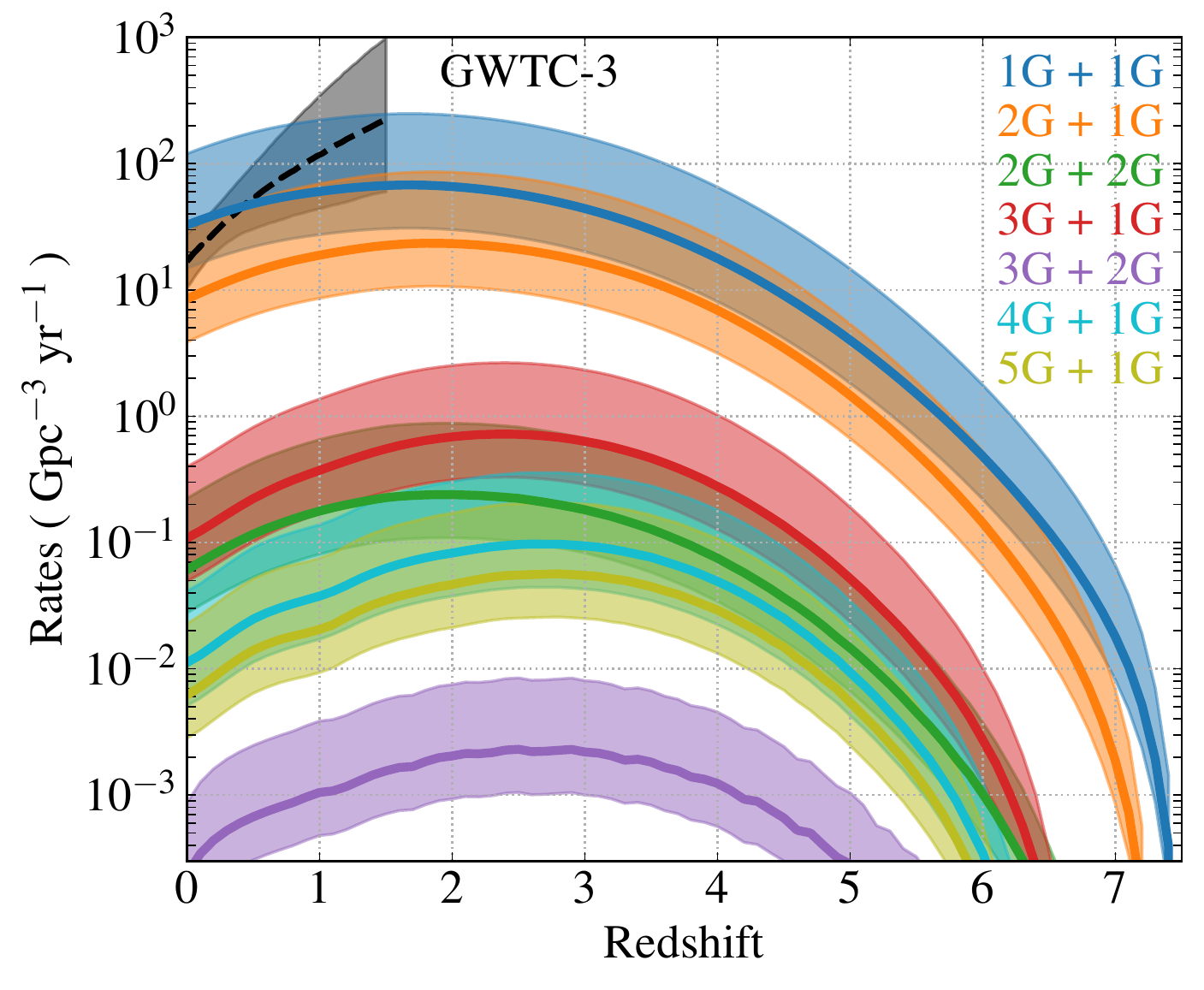}
\caption{Predicted merger rates for various generations of BBH mergers, assuming a cluster mass distribution $\propto M_{\rm CL}^{-2}$ up to a maximum mass of $10^7\msun$. Here the half-mass radius of all clusters is fixed at $1$\,pc. The black area represents the $90\%$ credible bounds on the BBH merger rate in the LVK analysis \citep{lvkcoll-O3-2}.}
\label{fig:rates_m7r1}
\end{figure}

\begin{figure} 
\centering
\includegraphics[scale=0.6]{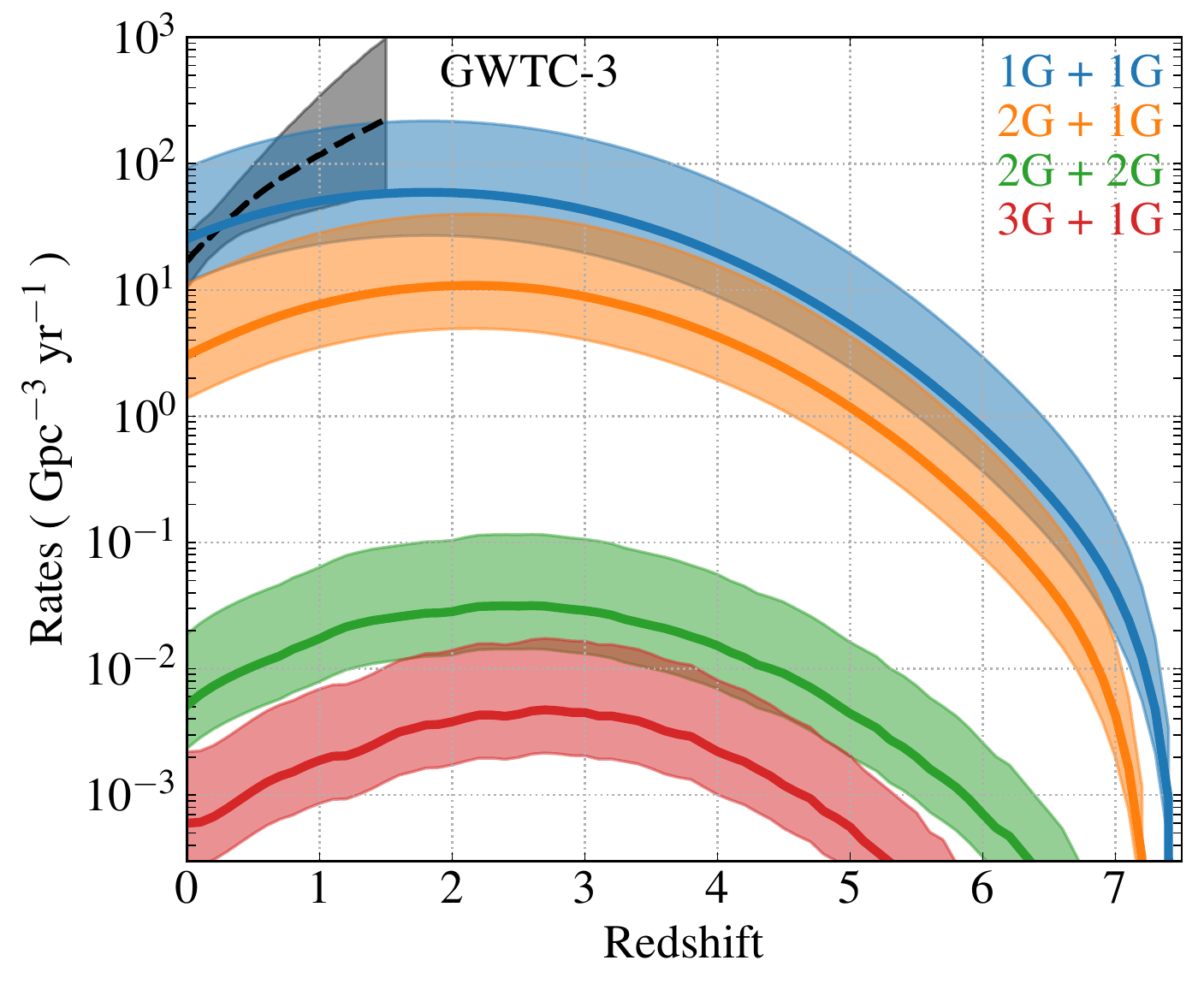}
\caption{Same as Figure~\ref{fig:rates_m7r1}, but with a maximum cluster mass lowered to $10^6\msun$.}
\label{fig:rates_m6r1}
\end{figure}

\begin{figure} 
\centering
\includegraphics[scale=0.6]{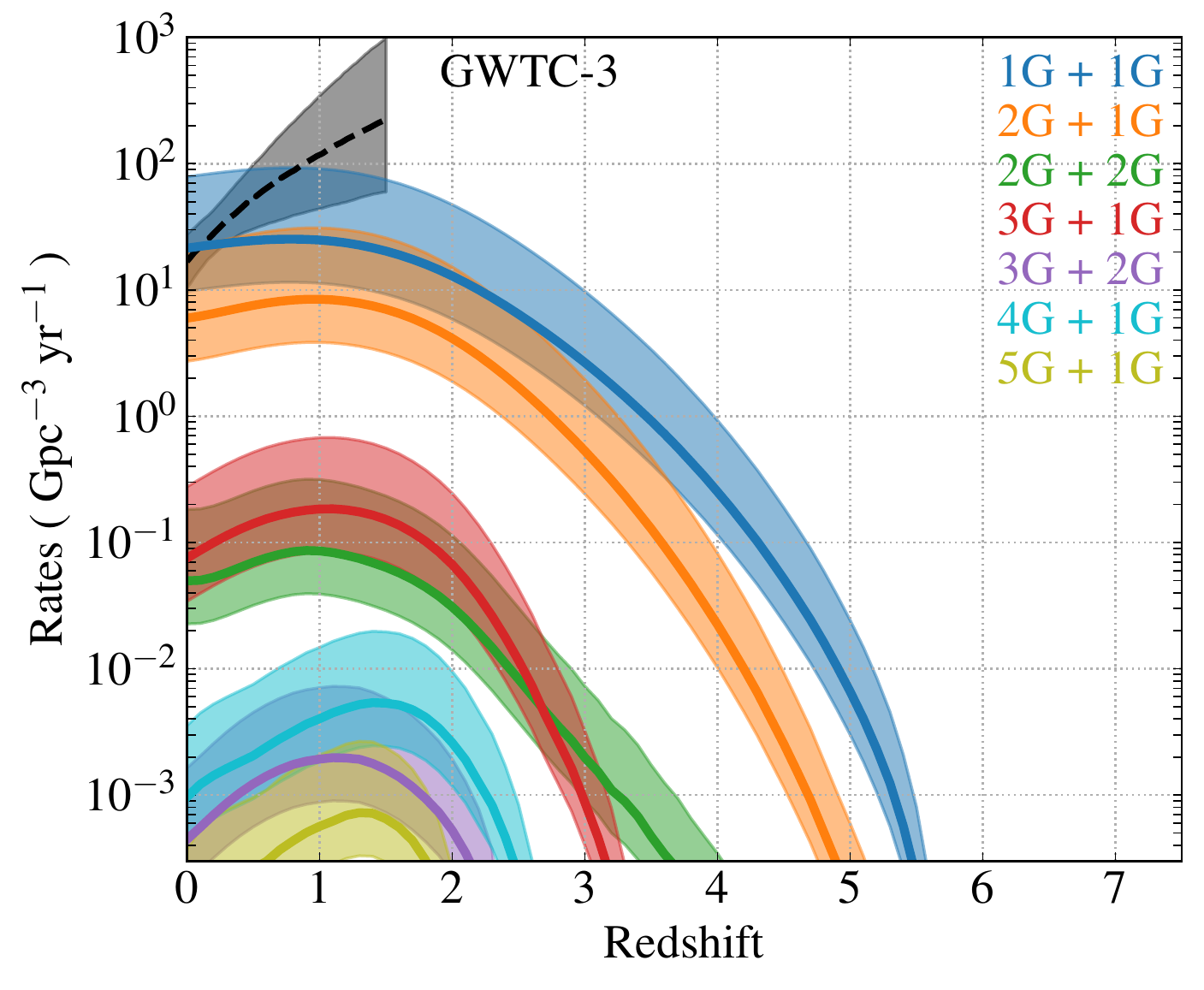}
\caption{Same as Figure~\ref{fig:rates_m7r1}, but with the half-mass radius of all clusters set to $3$\,pc.}
\label{fig:rates_m7r3}
\end{figure}

Figure~\ref{fig:rates_m7r1} shows the merger rates of various generations of BBH mergers, assuming a cluster mass distribution $\propto M_{\rm CL}^{-2}$ up to a maximum mass of $10^7\msun$. The half-mass radius of all clusters is fixed at $1$\,pc. In this case, our models predict a mean merger rate of about $30\gpcyr$ at $z=0$ for 1G+1G mergers, while this becomes about $8\gpcyr$, $1\times 10^{-1}\gpcyr$, $1\times 10^{-2}\gpcyr$, $7\times 10^{-3}\gpcyr$, $7\times 10^{-2}\gpcyr$, and $3\times 10^{-4}\gpcyr$ for 2G+1G, 3G+1G, 4G+1G, 5G+1G, 2G+2G and 3G+2G mergers, respectively. For reference, the LVK rate for BBH mergers is between $17.9\gpcyr$ and $44\gpcyr$ \citep{lvkcoll-O3-2}. When the star cluster mass distribution is truncated to a maximum mass of $10^6\msun$ (see Figure~\ref{fig:rates_m6r1}), we find that the mean rate of 1G+1G mergers slightly decreases, to about $25\gpcyr$ at $z=0$, while the merger of higher generations decreases more significantly. In particular, we find that 2G+1G, 2G+2G, and 3G+1G mergers have a merger rate of $3\gpcyr$, $5\times 10^{-3}\gpcyr$, and $5\times 10^{-4}\gpcyr$, respectively, with no merger seen on our models with fourth- or higher-generation BHs. This reflects the fact that in this case there are no massive star clusters ($\gtrsim 3\times 10^6\msun$, see Figure~\ref{fig:gener}) that can retain a 4G BH. Finally, we plot the merger rates of BBH mergers, assuming a cluster mass distribution $\propto M_{\rm CL}^{-2}$ up to a maximum mass of $10^7\msun$ and half-mass radius of all clusters fixed at the larger value of $3$\,pc in Figure~\ref{fig:rates_m7r3}. It is clear that the rate of 1G+1G, 2G+1G, 2G+2G, and 3G+1G mergers at $z=0$ does not significantly change with respect to the case of star clusters with smaller half-mass radii, while the merger rates for higher generations are smaller. Also the peak and the shape of the rate distributions as a function of redshift are affected by the initial choice of half-mass radius. This illustrates how detecting  hierarchical mergers could constrain the overall distributions of cluster masses and densities, which have an imprint on the rates of BBH mergers, and their evolution across cosmic time.

\begin{figure*} 
\centering
\includegraphics[scale=0.45]{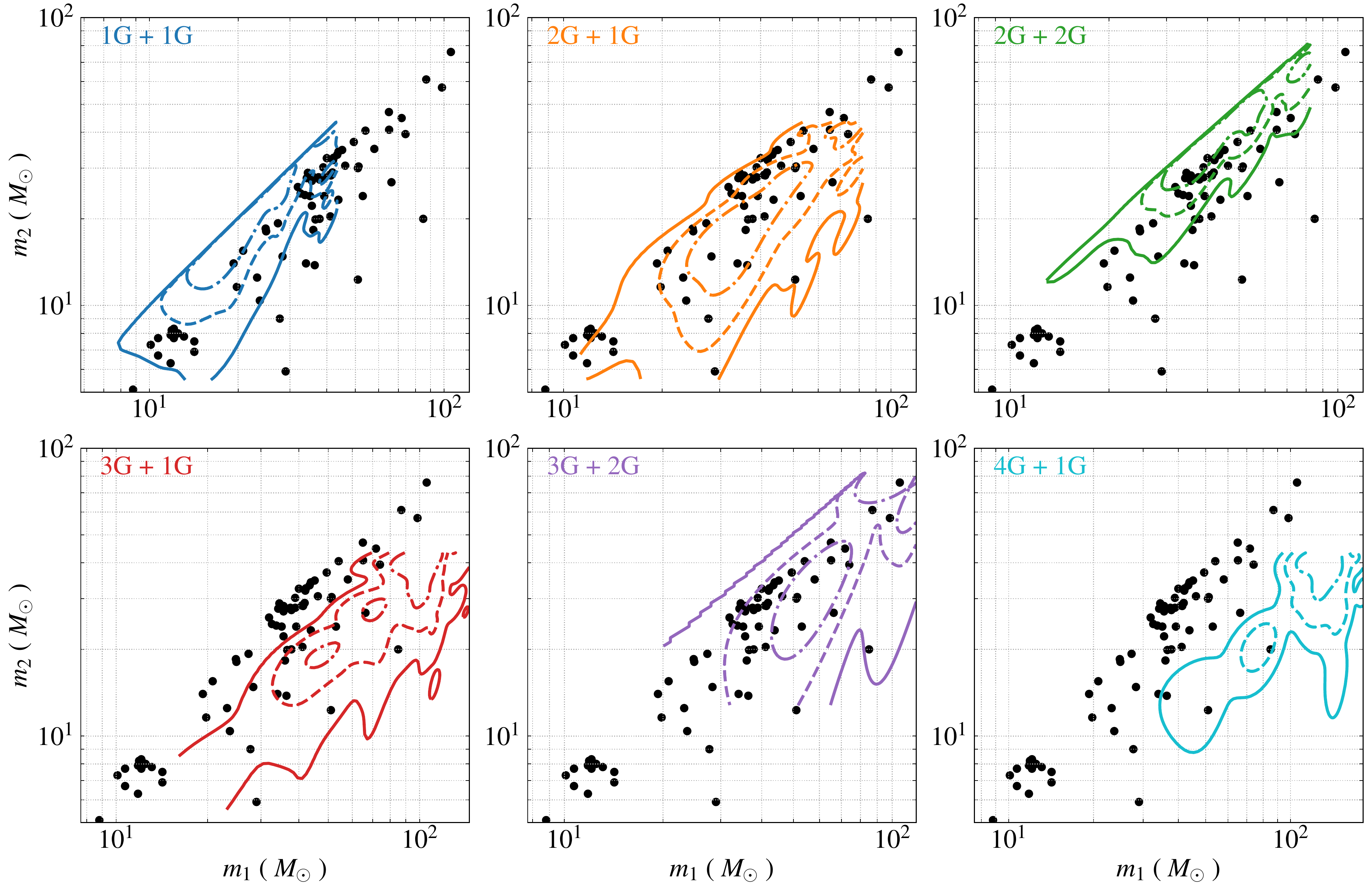}
\caption{Component masses of merging BBHs detected by the LVK Collaboration \citep{lvkcoll-O3} and in models assuming a cluster mass distribution $\propto M_{\rm CL}^{-2}$ up to a maximum mass of $10^7\msun$ and $r_{\rm h}=1$\,pc, for various generations of mergers. Dot-dashed, dashed, and solid lines represent the $1\sigma$, $2\sigma$, $3\sigma$ contours of the distributions, obtained by weighting the simulation results with the detection likelihood $w_{\rm det}$.}
\label{fig:ctp_lvk}
\end{figure*}

We note that our model predicts a 1G+1G merger rate that matches the LVK observed rate, given our particular choice of cluster parameters and mass distribution. Clearly, there are more than one astrophysical scenario that contribute to the overall observed population \citep[e.g.,][]{MandelBroekgaarden2022}. Indeed, the main goal of our study is not to reproduce exactly the observed LVK rates, which we leave to a further study, rather we want to show general trends in the BBH merger rate for first- and higher-generations and how the uncertain parameters of the distributions that describe star clusters across cosmic time affect them. For example, we have shown that bigger clusters tend to assemble more merging binaries and that less dense clusters produce fewer mergers, hence fewer repeated mergers. Moreover, we want to note that if star clusters are relatively dense, the runaway merger of main-sequence stars could happen, producing a very massive star that can possibly collapse to form an IMBH \citep[e.g.,][]{por02,gie15,GonzalezKremer2021}. This would affect the number of repeated mergers and the evolutionary pathways that shape the growth of an IMBH \citep[e.g.,][]{fragk18,fragleiginkoc18}.

We now compare the masses of the merging BBHs of different generations that we find in our simulations with the detected population by the LVK Collaboration \citep{lvkcoll-O3}. In order to do that, we start by accounting for the observational weights by advanced GW observatories, considering the increased sensitivity of the detectors to BBHs of higher masses and the larger amount of comoving volume surveyed at higher redshifts. In addition to the weights accounting for the distribution of masses, formation times, and metallicity of the parent dense star cluster, we assign each BBH a detectability weight defined as \cite[see, e.g.,][]{FragioneBanerjee2021}
\begin{equation}
w_{\rm det} = \frac{p_{\rm det}(m_1,m_2,z)}{1+z} \frac{dVc}{dz}\,,
\label{eqn:gwweight}
\end{equation}
where $dV_c/dz$ is the amount of co-moving volume in a slice of the universe at redshift $z$, $1/(1+z)$ is the difference in comoving time between the merger redshift and the observer at $z=0$, and $p_{\rm det}(m_1,m_2,z)$ is the detection probability of sources with masses $m_1$ and $m_2$ merging at redshift $z$. To compute the GW detectability signal-to-noise (S/N) ratio, we use the \textsc{IMRPhenomD} GW approximant \citep{SantamariaOhme2010} and assume a single LIGO instrument at design sensitivity, following the procedure outlined by \citet{DominikBelczynski2013}. We define the detection probability $p_{\rm det}(m_1,m_2,z)$ as the fraction of sources of a given mass located at the given redshift that exceed the detectability threshold in S/N, assuming that sources are uniformly and isotropically distributed in sky location and orbital orientation
\begin{equation}
p_{\rm det}(m_1,m_2,z)=P(\rho_{\rm thr}/\rho_{\rm opt})\,,
\label{eqn:detec}
\end{equation}
where $\rho_{\rm opt}$ is the S/N ratio for an optimally located and oriented (face-on and directly overhead) binary and $\rho_{\rm thr}=8$ is the S/N ratio threshold, and
\begin{eqnarray}
P(\mathcal{W})&=&a_2(1-\mathcal{W})^2+a_4(1-\mathcal{W})^4+a_8(1-\mathcal{W})^8\nonumber\\
&+&(1-a_2-a_4-a_8)(1-\mathcal{W})^{10}\,,
\end{eqnarray}
where $a_2=0.374222$, $a_4=2.04216$, and $a_8=-2.63948$.

Figure~\ref{fig:ctp_lvk} shows a comparison of component masses for merging BBHs detected by the LVK Collaboration \citep{lvkcoll-O3} and in our models, assuming a cluster mass distribution $\propto M_{\rm CL}^{-2}$ up to a maximum mass of $10^7\msun$ and half-mass radius $r_{\rm h}=1$\,pc, for various generations of mergers\footnote{This choice of the value of the initial half-mass radius is consistent with the mean value of $r_{\rm h}$ needed to reproduce the LVK rate for dynamical mergers \citep{FishbachFragione2023}.}. This plot shows that, within our models, some events can only be explained by higher BH generations. In particular, GW190521, GW190426\_190642, and GW200220\_061928 are consistent with coming from 3G+2G mergers. Besides the agreement in component masses, a full analysis of these signals, and the determination of which formation channel is most likely for each one, would also require careful consideration of the BH spins (see Figure~\ref{fig:spin}), which we leave to a future work.

\section{Discussion and conclusions}
\label{sect:concl}

Although the LVK collaboration has detected more than $80$ merging BBHs, the exact shape of the BH mass spectrum remains poorly known. Current stellar evolution models predict a dearth of BHs with masses $\gtrsim 50\msun$ as a result of pair-instability physics, but the detection of GW190521 and other events with one or both component masses above this limit has challenged theoretical models.

BHs with higher masses could be produced through repeated mergers of smaller BHs in the center of a dense star cluster. Here, the high stellar density in the core leads to efficient formation of merging BBHs, and provides a deep potential well that could retain merger remnants even when they receive a relativistic recoil kick of hundreds $\kms$. The merger remnant could then undergo the same dynamical processes and eventually merge with another BH via GW emission. The likelihood of this hierarchical merger process is very sensitive to the cluster mass and density: the higher the mass and density are, the more likely it is. Unfortunately, the most interesting star clusters cannot be simulated numerically with direct (Aarseth-type) $N$-body codes, and even parallel Monte Carlo codes remain limited in this regime of very large cluster masses with high densities.

In this paper, we have used a semi-analytic framework to investigate hierarchical mergers in dense star clusters, based on a method first developed by \citet{AntoniniGieles2020}. Our method allows us to rapidly study the outcomes of hierarchical mergers as a function of the cluster masses, densities and metallicities. We have discussed the characteristics of the population of higher-generation BHs and their GW signatures.

We have shown in some detail how the likelihood of higher-generation mergers increases with cluster mass and density. Assuming a half-mass radius of $1$\,pc, we have found that 1G+1G mergers represent most of the population of BBH mergers for clusters masses of $\sim 10^5\msun$, with 2G+1G mergers becoming $\sim 10\%$ of the population for clusters masses of $\sim 10^6\msun$,  and up to about $30\%$ for cluster masses around $5\times 10^6\msun$. Mergers of 3G+1G BBHs start happening for clusters masses of $\sim 10^6\msun$ and are typically never more than $\sim 1\%$ of all mergers, while 4G+1G mergers are assembled only for cluster masses $\sim 3\times 10^6\msun$. This trend is then essentially reproduced by any higher-generation merger (5G+1G, 6G+1G, and so). The reason for this is that the mass ratio of the merger starts becoming quite small and the recoil kick imparted to the remnant is no longer large enough to eject it from the parent cluster. Around $4\times 10^6\msun$, a single BH starts to dominate the mergers and can grow all the way to the IMBH regime, $\gtrsim 1000\msun$. We have also shown that the overall trends mainly depend on the initial cluster mass and radius, and not on its metallicity.

Assuming a cluster mass distribution $\propto M_{\rm CL}^{-2}$ up to a maximum mass of $10^7\msun$ and half-mass radius of the clusters fixed to $1$\,pc, our models predict a mean merger rate of about $30\gpcyr$ at $z=0$ for 1G+1G mergers, and about $8\gpcyr$, $1\times 10^{-1}\gpcyr$, $1\times 10^{-2}\gpcyr$, $7\times 10^{-3}\gpcyr$, $7\times 10^{-2}\gpcyr$, and $3\times 10^{-4}\gpcyr$ for 2G+1G, 3G+1G, 4G+1G, 5G+1G, 2G+2G and 3G+2G mergers, respectively. If the star cluster mass distribution is instead truncated at a maximum mass of $10^6\msun$ or if we assume a larger initial half-mass radius of $3$\,pc, we have found that the rate of 1G+1G mergers slightly decreases, to about $25\gpcyr$ at $z=0$, while for higher generations the rates decrease more significantly. The location of the peak and the overall shape of the rates as a function of redshift are also affected by the initial choice of half-mass radius. 

Finally, we have discussed the few detected GW sources that can only be explained by higher BH generations. In particular, GW190521, GW190426\_190642, and GW200220\_061928 are consistent with being 3G+2G mergers. Our results can be used to inform detailed Bayesian inference to assess the likelihood of detected events of being consistent with higher-generation mergers, based on their masses, mass ratios, and effective spins \citep[e.g.,][]{KimballTalbot2021}. We leave such a detailed study to future work.

While we refer the reader to \citet{AntoniniGieles2020} for a full discussion of the uncertainties in our simplified cluster models, we want to point out that we do not model the effect of external tidal fields. For example, clusters with stronger tidal fields would be typically more compact, which might favor BBH mergers, but are also more susceptible to tidal disruption. We also do not account for primordial binary stars. Some fraction of them could become BBHs, which would then be an essential ingredient in the early dynamical evolution of the star cluster. After the segregation of BHs, the central energy generation will be shared by dynamically-assembled binaries and primordial binaries, potentially affecting the relevant encounter rates and BH mergers. However, we do not expect primordial binaries to have a significant effect on the overall rates at lower redshift once they have been dynamically processed, eventually merging or exchanging one of their components.

We also want to stress that one of the main sources of uncertainty in predicted merger rates for BBHs is the poorly known distributions of cluster properties (masses, radii, metallicities, and formation times) across the Universe. While most of these distributions are difficult to determine observationally \citep[for a review see][]{PortegiesZwartMcMillan2010}, some of them may soon be constrained directly by JWST observations \citep[e.g.,][]{MowlaIyer2022,VanzellaClaeyssens2022}. On the other hand, the current and upcoming detections of GW sources can be used to constrain them indirectly, assuming that some fraction of the population is indeed assembled dynamically in dense star clusters \citep{FishbachFragione2023}. Importantly, when all these considerations are taken carefully into account, dense star clusters may be found to produce a majority of detectable BBH mergers.

With the start of the next LVK run, hundreds of additional BBH mergers are expected to be detected over the next few years. Assuming $\sim 500$ BBH mergers detected in O4 by LVK, we predict that $\sim 50$ and $\sim 5$ of these events will contain a 2G and 3G BH, respectively, and up to 1~event could involve a 4G BH. With their distinctive signatures of higher masses and spins, hierarchical mergers offer an unprecedented opportunity to learn about dense star clusters throughout the Universe and to shed light on the elusive population of IMBHs. 

\section*{Acknowledgements}

This work was supported by NASA Grant 80NSSC21K1722 and NSF Grant AST-2108624 at Northwestern University.

\appendix
\restartappendixnumbering

\section{Comparison between semi-analytical framework and Cluster Monte Carlo}
\label{app:comparison}

Here, we compare the results of our semi-analytical method to the results obtained in detailed Monte Carlo simulations of GCs. In particular, we compare the number of GW-capture, in-cluster, and ejected BBH mergers from our models to the \textsc{cmc} Cluster Catalog \citep{KremerYe2020}. This catalog of models was obtained using the publicly available code \textsc{cmc} \citep{RodriguezWeatherford2022}, which  incorporates all the relevant physics for the evolution of dense star clusters, including two-body relaxation, three-body binary formation, strong three- and four-body interactions, some post-Newtonian effects, stellar evolution of single stars and binary stars, respectively. The catalog spans a wide range of initial conditions, including different initial numbers of stars ($N = 2\times10^5$, $4\times10^5$, $8\times10^5$, $1.6\times10^6$), corresponding to stellar masses ($M/\msun =1.2\times10^5$, $2.4\times10^5$, $4.8\times10^5$, $9.6\times10^6$), virial radii ($r_{\rm v}/\rm{pc} = 0.5, 1, 2, 4$), metallicities ($Z = 0.0002, 0.002, 0.02$), and Galactocentric distances ($R_{\rm g}/\rm{kpc}=2, 8, 20$).

\begin{figure*} 
\centering
\includegraphics[scale=0.525]{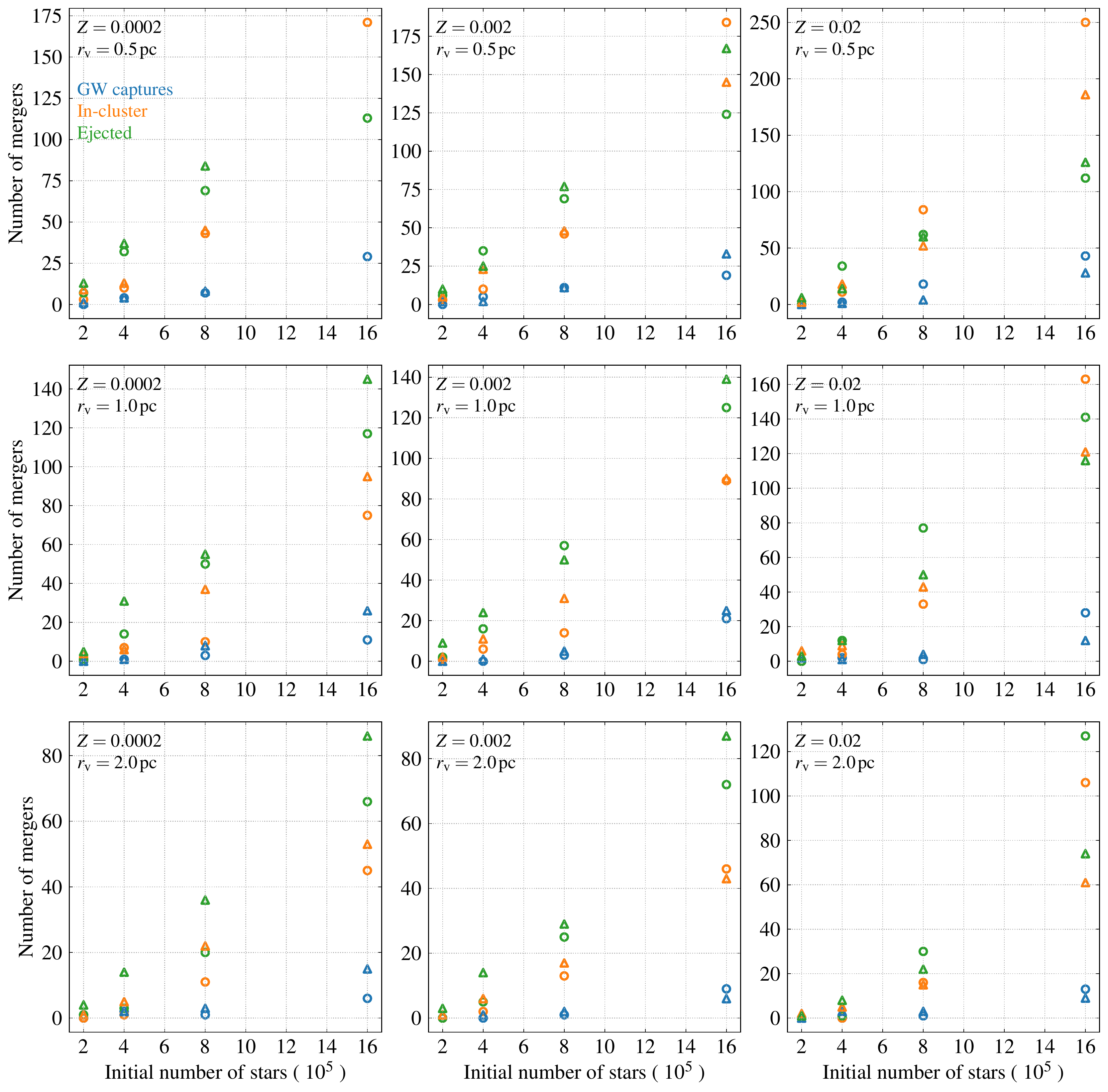}
\caption{Number of GW-capture, in-cluster, and ejected mergers in the models with Galactocentric distance $20$\,kpc in the \textsc{cmc} Cluster Catalog \citep[triangles;][]{KremerYe2020} and using our semi-analytical method (circles; see Sect.~\ref{sect:method}) as a function of the initial number of stars. Different panels show different metallicities (left: $Z=0.0002$; center: $Z=0.002$; right: $Z=0.02$) and virial radii (top $r_{\rm v}=0.5$\,pc; center $r_{\rm v}=1.0$\,pc; bottom $r_{\rm v}=2.0$\,pc). Note that there is no \textsc{cmc} model with initial number of stars $1.6\times 10^6\msun$ and $(Z, r_{\rm v})=(0.0002, 0.5\,{\rm pc})$.}
\label{fig:comp}
\end{figure*}

\begin{figure*} 
\centering
\includegraphics[scale=0.525]{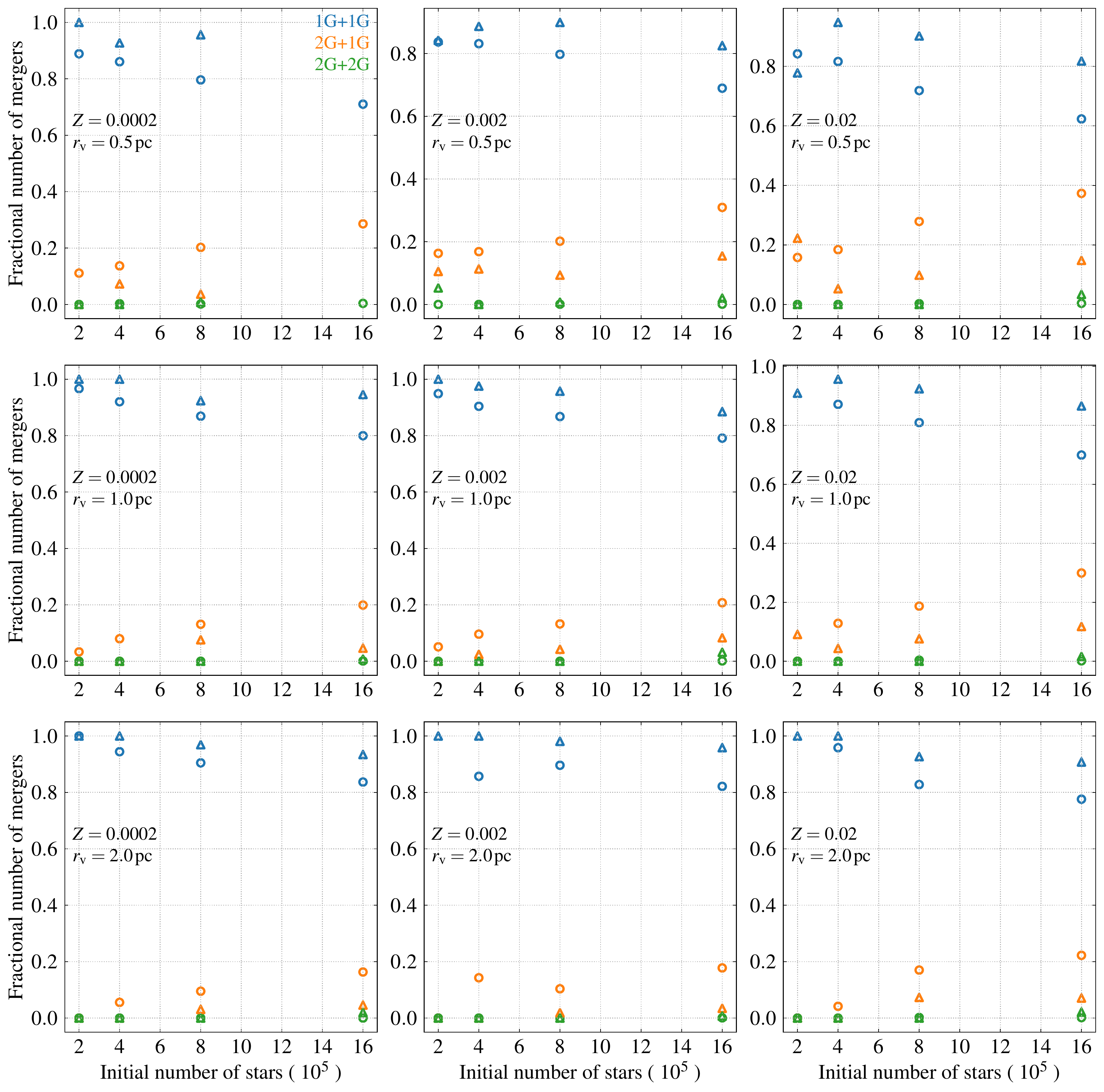}
\caption{Same as Figure~\ref{fig:comp}, but for the fractional number of 1G+1G, 2G+1G, and 2G+2G mergers.}
\label{fig:comp2g}
\end{figure*}

Figure~\ref{fig:comp} shows the number of GW-capture, in-cluster, and ejected mergers in the \textsc{cmc} Cluster Catalog and using our semi-analytical method (avereged over $10$ realizations for each combination of initial number of stars, virial radius, and metallicity) for different initial number of stars, metallicities, and virial radii; Figure~\ref{fig:comp2g} shows the same comparison for the fractional number of 1G+1G, 2G+1G, and 2G+2G mergers. Note that for our adopted models $r_{\rm h}\approx (3/4) r_{\rm v}$. We just use the models in the \textsc{cmc} Cluster Catalog that have Galactocentric distance of $20$\,kpc since our semi-analytical treatment does not include prescriptions for tidal stripping of stars. We find that the branching ratios for different BBH mergers and their overall normalization are quite fairly reproduced, given the approximate nature of our semi-analytical treatment of cluster and BHs evolution \citep[see also][]{AntoniniGieles2020}. The agreement is a result of the balanced evolution between the host star cluster and its BH population, which dictates the properties of BBH mergers.

\bibliography{refs}

\end{document}